\documentclass[12pt]{article}

\usepackage{amsmath, bm, amssymb, amsthm}
\usepackage{array}
\usepackage{comment}
\usepackage{dsfont}
\usepackage{enumerate}
\usepackage[margin=1in]{geometry}
\usepackage{graphicx}
\usepackage{hyperref}
\usepackage{latexsym}
\usepackage{lineno}
\usepackage{longtable, float}
\usepackage{lscape}
\usepackage{mathtools}
\usepackage{multirow}
\usepackage{rotating}
\usepackage[onehalfspacing]{setspace}
\usepackage{subcaption}
\usepackage{tabularx}
\usepackage{thmtools}
\usepackage{threeparttablex}
\usepackage{threeparttable}
\usepackage{xcolor}

\usepackage{booktabs}
\usepackage{colortbl}
\usepackage{makecell}
\usepackage{pdflscape}
\usepackage{tabu}
\usepackage[normalem]{ulem}
\usepackage{wrapfig}

\newtheorem{theorem}{Theorem}[section]
\newtheorem{prop}{Proposition}[section]
\newtheorem{corollary}[theorem]{Corollary}
\newtheorem{lemma}[theorem]{Lemma}
\theoremstyle{definition}

\theoremstyle{definition}
\newtheorem{assumption}{Assumption}[section]
\theoremstyle{definition}

\theoremstyle{remark}

\theoremstyle{remark}

\DeclareMathOperator{\EX}{\mathbb{E}}
\newcommand{\norm}[1]{\left\lVert#1\right\rVert}
\DeclareMathOperator{\Real}{Re}
\DeclareMathOperator{\Imag}{Im}

\usepackage{pgfplots}
\pgfplotsset{width=10cm,compat=1.9}
\usepgfplotslibrary{external}
\usetikzlibrary{through,calc,arrows,intersections,patterns}
\usetikzlibrary{decorations.pathreplacing}
\usetikzlibrary{decorations.markings}
	\tikzset{->-/.style={decoration={markings,mark=at position #1 with {\arrow{>}}},postaction={decorate}}}

\usepackage[round,authoryear]{natbib}
\title{Testing for Idiosyncratic Treatment Effect Heterogeneity\footnote{I thank Dick Startz, Doug Steigerwald, Cl\'ement de Chaisemartin, Pedro Sant'Anna, Gonzalo Vazquez-Bare, Ivan Canay, Yingying Dong, Shuo Qi, and the assistants of the Midwest Economics Association 2023, RCEA Econometrics Conference 2022, Microsoft Research Seminar, Australasia Meetings of the Econometric Society 2022 for valuable comments in earlier versions of this manuscript. Use was made of computational facilities purchased with funds from the National Science
Foundation (CNS-1725797) and administered by the Center for Scientific Computing (CSC). The CSC is supported
by the California NanoSystems Institute and the Materials Research Science and Engineering Center (MRSEC; NSF
DMR 1720256) at UC Santa Barbara.}}
\author{Jaime Ramirez-Cuellar\footnote{Microsoft, Office of the Chief Economist. Address: 14820 NE 36th Street, Redmond, Washington, 98052. Email: \texttt{jrc@econ.ucsb.edu}.}}
\date{\today}

\begin{document}

\maketitle

\begin{abstract}
    This paper provides asymptotically valid tests for the null hypothesis of no treatment effect heterogeneity. Importantly, I consider the presence of heterogeneity that is not explained by observed characteristics, or so-called idiosyncratic heterogeneity. When examining this heterogeneity, common statistical tests encounter a nuisance parameter problem in the average treatment effect which renders the asymptotic distribution of the test statistic dependent on that parameter. I propose an asymptotically valid test that circumvents the estimation of that parameter using the empirical characteristic function. A simulation study illustrates not only the test's validity but its higher power in rejecting a false null as compared to current tests. Furthermore, I show the method's usefulness through its application to a microfinance experiment in Bosnia and Herzegovina. In this experiment and for outcomes related to loan take-up and self-employment, the tests suggest that treatment effect heterogeneity does not seem to be completely accounted for by baseline characteristics. For those outcomes, researchers could potentially try to collect more baseline characteristics to inspect the remaining treatment effect heterogeneity, and potentially, improve treatment targeting. 
\end{abstract}

\newpage

\section{Introduction}

The literature on program evaluation interrogates whether and how the effect of a program varies across people. The literature argues that such investigation helps assess important causal mechanisms for treatment effect heterogeneity based on observed characteristics \citep{crump2008nonparametric,ding2019decomposing}. Likewise, policymakers often wish to discern who benefits the most out of a policy intervention, to administer the intervention to those individuals, and to exclude other individuals, so-called targeting \citep{manski2004statistical,dehejia2005program,hirano2009asymptotics,bhattacharya2012inferring,kitagawa2018should,wager2018estimation}. After conducting such an exercise, it may be useful to know if there is still some treatment effect heterogeneity left, also known as idiosyncratic heterogeneity, across individuals with the same characteristics. If so, that may indicate that there is room for further targeting, and that collecting more individual characteristics could help improve the treatment assignment even further. It is therefore critical to establish whether there is unexplained, idiosyncratic heterogeneity in treatment effects. \citep{heckman1997making,djebbari2008heterogeneous,ding2016randomization}. 

This paper provides asymptotically valid tests for treatment effect heterogeneity. In deriving tests to detect idiosyncratic heterogeneity, I first explore tests to detect general heterogeneity. When analyzing general heterogeneity, this paper proposes a test that circumvents the estimation of the ATE using the characteristic function, which, unlike previous tests, renders the asymptotic distribution independent from the ATE.  Simulations show that the test detect true heterogeneity at higher rates than previous tests in small samples.

Additionally, a second test aims to detect differences across treatment and control groups after accounting for differences in their conditional means given covariates. Consequently, I propose a permutation procedure based on the characteristic function that is both asymptotically valid and consistent against general alternatives. Finally, I apply the methods to a microcredit experiment in Bosnia and Herzegovina aimed at marginally rejected loan applicants. 

I start with the simple hypothesis of no treatment effect heterogeneity. In this case, the null hypothesis implies that the distribution of potential outcomes for the treated and control groups shifts by a constant, which is the average treatment effect. A typical solution to this inference problem requires the researcher to estimate this constant, which acts as a parameter of no direct relevance to the problem. This parameter renders the asymptotic distribution of the test statistic dependent of the average treatment effect, which is often called the nuisance parameter or Durbin problem \citep{durbin1973distribution,basu1977elimination,sen1984aligned,babu2004goodness,ding2016randomization,CHUNG2021148}. To circumvent the estimation of this constant, I formulate a nonparametric test based on the symmetrized outcomes, which are the difference between two independent observations of the outcomes. Since the distribution of the symmetrized outcomes should be equivalent across the treatment and control groups under the null hypothesis, their characteristic functions should also be equivalent. Accordingly, I construct an asymptotically valid test that compares the characteristic functions of the symmetrized outcomes of the treatment and control groups.

I then introduce a procedure for the more prominent case of detecting idiosyncratic treatment effect heterogeneity or, equivalently, heterogeneity that is not explained by observed characteristics. First, I remove the variation in the observed outcomes predicted by covariates in separate regressions, by treatment and control groups, of the observed outcome on covariates. Using the residuals from those regressions, I perform a homogeneity test based on the empirical characteristic function to check for equality of the unexplained variation across treatment and control groups. I show that the test is asymptotically valid and consistent for general alternatives. Except for the use of regression residuals, the test statistic is identical to the one used in \citet{meintanis2005permutation} in the case with covariates. Given that the asymptotic distribution of the test statistic depends on the potential outcomes distributions, I implement a permutation test to approximate the asymptotic distribution and obtain critical values. This permutation test offers an alternative to current approximations based on the weighted bootstrap in the empirical characteristic function literature \citep{rivas2019two}. 

This paper contributes to the growing literature that tests for treatment effect variation in program evaluation. In the absence of covariates, \citet{ding2016randomization} use randomization inference and Kolmogorov-Smirnov statistics to test for idiosyncratic treatment effect variation, but their test requires the estimation of a confidence interval for the average treatment effect. %
\cite{sen1984aligned} suggests a test that uses the empirical distribution function under a restrictive symmetry assumption. In the case of no covariates, one can also compare the variances of the treated and untreated outcome distributions \citep{cox1984interaction,white1980heteroskedasticity,breusch1979simple} or compare other moments of these two counterfactual distributions. However, those tests can only be applied to distributions with finite moments, and still they encounter the nuisance parameter problem \citep{ding2016randomization}.\footnote{Other literature has explored systematic treatment effect variation explained by covariates in the form of quantile regression \citep{koenker2002inference,chernozhukov2005subsampling}, finite-population regressions \citep{ding2019decomposing}, nonparametric tests for equality of conditional average treatment effects \citep{crump2008nonparametric}, multiple testing for treatment effect heterogeneity across subgroups defined by discrete covariates \citep{lee2014multiple}, variation in subgroups' average treatment effects explaining treatment effect heterogeneity \citep{bitler2017can,buhl2022some}, censored outcomes \citep{sant2016nonparametric}, and multiple testing comparison of quantiles across treatment and control groups \citep{goldman2018comparing}.} Likewise, in the case without covariates, \citet{CHUNG2021148} propose a permutation test that eliminates the nuisance parameter using a martingale transformation of the empirical process; however, unlike the test based on the characteristic function, that test requires the outcomes to be continuous.

The use of the empirical characteristic function for hypothesis testing is not new. To construct such statistical tests, \citet{feuerverger1977empirical} and \citet{csorgo1981limit} show point-wise consistency of the empirical characteristic function to its population counterpart, which is implied by a mild moment condition. Using the characteristic function, \citet{meintanis2005permutation} and \citet{chen2019some} introduce tests for equality of two distributions, joint independence and univariate symmetry. In the case of no covariates available to the researcher, the test based on the characteristic function is closest to the location-shift test in \citet{henze2005checking}, but that paper requires the estimation of the nuisance parameter. In contrast, I avoid the estimation of the true treatment effect using a statistic that does not depend on this parameter.\footnote{The statistic in this paper is equivalent to the the difference in weighted integrals of the characteristic function's squared modulus of the treated and control groups. Several authors have used the squared modulus of the characteristic function \citep[for a review of older literature, see][]{meintanis1996robustness}. For instance, using this statistic, \citet{hu2021simple} test for completeness in a class of nonparametric models, and \citet{murota1981studentized} formulate a normality test based on standardization of the empirical characteristic function.}
%

In a simulation study, I compare the small sample properties of the tests to competing methods based on the characteristic function, the distribution function (specifically, the Kolmogorov-Smirnov tests), and the quantile regression process. In the case of general heterogeneity, the tests proposed in this paper achieve higher power compared to the methods based on the characteristic function  \citep{henze2005checking}, the confidence interval randomization inference test \citep{ding2016randomization} and the subsampling test based on the the quantile regression process \citep{chernozhukov2005subsampling}. I also compare two versions of this paper's tests: one that circumvents the estimation of the average treatment effect and one that does not. Accordingly, the simulations show that circumventing the estimation of the average treatment effect has zero or at most modest gains in small samples. This modest gains are likely due to two main reasons: the success of resampling techniques in approximating the asymptotic distribution of the test statistic without worrying about the effect of unknown parameters such as the average treatment effect or the potential outcomes distributions \citep{bickel1969distribution,romano1988bootstrap,praestgaard1995permutation,abadie2002bootstrap}, and secondly, the consistency of the common difference-in-means estimator. I also apply the confidence interval randomization test in \citet{berger1994p} and \citet{ding2016randomization} to the tests based on the characteristic function. In the simulations, I find that the confidence interval randomization test is conservative and loses power compared to the procedure I propose and that of \citet{henze2005checking}. 

In the more salient case of covariates available to the researcher, the simulations show the validity and power of the homogeneity test that compares the characteristic function of the observed outcomes' residuals. Given the relative success of the resampling techniques for the case without covariates, I propose a permutation test that has power against alternatives.

I apply the characteristic function tests to 40 outcomes in the microfinance experiment that \citet{augsburg2015impacts} implemented in Bosnia and Herzegovina. That experiment targeted 1,196 loan applicants who were marginally rejected by a microfinance institution, with 628 of these applicants randomly selected for a typical loan. As one would expect most of the outcomes for which the authors found significant treatment effects are the candidates to exhibit treatment effect variation. Among these outcomes, there are several outcomes for which the covariates collected by the authors do not completely explain treatment effect variation including loan take-up, ownership of inventory and hours worked by working-age teenagers in the household business. For those outcomes, researchers could potentially try to collect more covariates to inspect the remaining treatment effect heterogeneity. For four outcomes, the tests suggest that any treatment effect variation seems to be accounted for by the covariates collected by the researchers; these outcomes include the incidence of self-employment income and the incidence of wage income, main business expenses and hours worked in activities other than business and wage labor. Finally, the test based on the statistic in \citet{henze2005checking} tends to reject more often than the tests with and without covariates proposed in this paper.  

The paper proceeds as follows. The second section introduces the setup. The third section explains the test based on the empirical characteristic function. The fourth section shows simulations exploring the small sample properties of the tests. The fifth section presents the application. The sixth section concludes. All proofs are included in the Appendix.



\section{Setup}

Consider a sample of $n$ units indexed by $i=1,\dots,n$. Let $D_i$ be a binary variable that indicates treatment status of individual $i$: $D_i=1$ if individual $i$ was treated, and $D_i=0$ otherwise. Let $Y_i(1)$ and $Y_i(0)$ represent random potential outcomes of individual $i$ with and without treatment, respectively. The observed outcome is given by
\begin{gather}
    Y_i = Y_i(1)D_i +Y_i(0)(1-D_i),    \qquad i=1,\dots,n.
    \label{eq:obs}
\end{gather}
Let $n_1$ and $n_0=n-n_1$ be the number of treated and control units, respectively. Let $\mathbf{X}_i$ be a $(q\times 1)$ vector of covariates not affected by treatment for unit $i$, which may include a constant, and let  $\mathbf{X}_i'$ be its transpose.

I now make two assumptions on the treatment assignment and sampling mechanisms. 
\begin{assumption}[Random assignment]
\label{asm:random_assign}
The assignment mechanism is such that \linebreak $[Y_1(0),Y_1(1)],\dots,[Y_n(0),Y_n(1)]$ and  $D_1,\dots,D_n$ are independent.
\end{assumption}
\begin{assumption} [Random sampling]
\label{asm:asm_randsample}
Let $[Y_{1}(0),Y_{1}(1), \mathbf{X}_1],\dots,[Y_{n}(0),Y_{n}(1), \mathbf{X}_n]$ be a random sample from some common distribution $F$.
\end{assumption}

Assumption \ref{asm:random_assign} guarantees that the treatment status does not influence the potential outcomes. Assumption \ref{asm:asm_randsample} states that the researcher has a random sample of units from the same population, which is a commonly made assumption in the literature \citep{crump2008nonparametric,wager2018estimation,heckman1997making,djebbari2008heterogeneous}.

This paper aims to test for equality in the distributions of the treated and control potential outcomes after taking into account different conditional mean responses given covariates. For instance, if gender is a covariate of interest, the paper aims to  answer whether there are differences between the distributions of the treatment and control groups after allowing for different mean responses for female and non-female participants by treatment assignment. This hypothesis could be of importance as a participant would be expected to have the same uncertainty (i.e., having the same distribution) over their response in either the treatment or the control group once the mean response specific by gender and treatment assignment is removed.\footnote{The paper differentiate itself from the literature that test for constant conditional average treatment effect \citet{crump2008nonparametric}. Likewise, the paper does not pursue any decomposition or inference over the treatment effect, $Y_i(1)-Y_i(0)$, which is in general an unobservable quantity \citep[see][]{ding2019decomposing}. } 

To make this purpose explicit, one can write the following decomposition of the potential outcomes into systematic and idiosyncratic components
\begin{align}
    \begin{split}
    Y(1) &=  \EX[Y(1)|\mathbf{X}]+\varepsilon(1) \text{ and }
    \\
    Y(0) &=  \EX[Y(0)|\mathbf{X}]+\varepsilon(0).
    \end{split}
\end{align}
The component captured by $\EX[Y(1)|\mathbf{X}]$ is the conditional expectation of the outcome when treated given covariates $\mathbf{X}$ or systematic component. The component given by $\varepsilon(1)$ is the part of the potential outcome after stripping down the systematic component. For instance, if $\mathbf{X}$ is a categorical variable that takes the values female and non-female, $\varepsilon(1)$ will be the treated potential outcome after removing the outcome means for female and non-female subpopulations.\footnote{\citet{ding2016randomization} used a similar procedure in a finite population setting where, instead of using the conditional expectation, they suggest using a linear regression projection $\mathbf{X}'\boldsymbol{\beta}$.}

The null hypothesis can be written as follows,
\begin{gather}
    \label{eq:h0}
    H_0:  \varepsilon(1) \overset{d}{=} \varepsilon(0),
\end{gather}
where $\overset{d}{=}$ means equality in distribution. This hypothesis conveys that the marginal distributions of treatment and control groups are the same after subtracting different conditional means given covariates.

To give simple intuition about $H_0$, in the simplest case of no covariates $\EX[Y(1)|\mathbf{X}]=\EX[Y(1)]$ and $\EX[Y(0)|\mathbf{X}]=\EX[Y(0)]$, so $H_0$ implies the following null hypothesis %
\begin{gather}
    H_{0}^\text{nocov}: Y(1) \overset{d}{=} Y(0)+\tau,
    \label{eq:h0noCov}
\end{gather}
where $\tau=\EX[Y(1)-Y(0)]$ is a fixed constant.

In this case, $H_{0}^\text{nocov}$ implies that the cumulative distribution functions of the potential outcomes when treated and control are shifted by a constant, $\tau$. Previous statistical tests require eliminating this constant from the statistical problem and multiple methods have been proposed such as studentization, substitution by estimated values, invariance principles, conditioning or maximizing across the parameter space, among several others \citep{basu1977elimination,lancaster2000incidental,koenker2002inference,henze2005checking,ding2016randomization,CHUNG2021148}.\footnote{Other alternatives include, as \citet{koenker2002inference} state, the use of resampling of the test statistic under conditions consistent with the null hypothesis in tests based on the empirical process  to obtain critical values \citep[see][]{bickel1969distribution,romano1988bootstrap,praestgaard1995permutation,abadie2002bootstrap}.} In the statistics literature, $H_0^\text{nocov}$ is also known as the location-shift hypothesis \citep[see][]{sen1984aligned}. \citet{cox1984interaction} asserts that if the marginal distributions of the treatment group stochastically dominates the marginal distribution of the control group, or vice versa, then there is a increasing transformation of the outcomes such as there is constant treatment effect in the transformed outcomes.


\section{A test for idiosyncratic treatment effect heterogeneity using the characteristic function}

\label{sec:test}

In this section, I introduce a hypothesis test using the characteristic function. To formulate a test for the null of no idiosyncratic treatment effect, I exploit the fact that the characteristic function has a one-to-one correspondence with the distribution function. I start with the simplest case when the researcher does not have access to exogenous characteristics $\mathbf{X}_i$, and I then develop the case when those variables are available.

\subsection{The no-covariates case}

To simplify the exposition, I first present the case in which the researcher is interested in testing for idiosyncratic treatment effect variation but there are no observed characteristics available. As noted in Equation \eqref{eq:h0noCov}, $H_0$ reduces to $H_{0}^\text{nocov}: Y(1) \overset{d}{=} Y(0) + \tau, \text{ for some } \tau \in \mathbb{R}.$

I now introduce a test based on the characteristic function to test for the null hypothesis $H_{0}^\text{nocov}$. Let $$\varphi_{d}(t)=\EX[e^{\mathrm{i}Y(d)t}]= \EX[\cos\{tY(d)\}+\mathrm{i}\sin\{tY(d)\}],$$ be the characteristic function of the potential outcome $Y(d)$ with treatment assignment $d$, for $d=0,1$, and $\mathrm{i}=\sqrt{-1}$. As noted in \citet{henze2005checking}, the null hypothesis $H_{0}^\text{nocov}$ implies that 
\begin{gather}
    \varphi_1(t) = e^{\mathrm{i}\tau t} \varphi_0(t), \quad \text{for all  $t \in \mathbb{R}$ and some $\tau \in \mathbb{R}$},
    \label{h0:noCovCF}
\end{gather}

After straightforward algebra and using the properties of the characteristic function, Equation \eqref{h0:noCovCF} in turn implies that
\begin{align}
    \varphi_1(t)\varphi_1(-t) 
    & = e^{\mathrm{i}\tau t}e^{-\mathrm{i}\tau t} \varphi_0(t) \varphi_0(-t),
    \notag
    \\
    & = \varphi_0(t) \varphi_0(-t),
    \label{eq:mult_phi}
\end{align}
for all  $t \in \mathbb{R}$. Then, one can restate Equation \eqref{eq:mult_phi} as
\begin{gather}
    |\varphi_1(t)|^2 = |\varphi_0(t)|^2 \text{ for all $t \in \mathbb{R}$},
    \label{eq:abs_varphi}
\end{gather}

where I use the fact that the modulus of the complex number $|\varphi_d(t)|^2$ is defined as $|\varphi_d(t)|^2=\varphi_d(t)\varphi_d(-t)$, $d=0,1$.\footnote{If $c$ is a complex number, then $|c|^2$ is the square modulus of $c$. Other tests using the square modulus of the characteristic function include, for example,  \citet{meintanis1996robustness}, \citet{murota1981studentized} and \citet{hu2021simple}.}

Note that Equation \eqref{eq:abs_varphi} does not depend on the average treatment effect $\tau$. To provide more intuition about this equation, for $d=0,1$, $|\varphi_1(d)|^2$ is also known as the characteristic function of the symmetrized variable $Y(d)-Y'(d)$, where $Y'(d)$ is independent from and has the same distribution as $Y(d)$ \citep[see][Corollary to Lemma XV.2]{feller2008introduction}. The connection with the symmetrized outcomes is indeed distinctive; symmetrized outcomes are centered at zero which implies that their associated location parameter or expected value is, if any, zero. In a nutshell, Equation \eqref{eq:abs_varphi} implies that the characteristic functions of the symmetrized outcomes of treatment and control groups is the same, and provides the advantage of removing the dependence on the average treatment effect, $\tau$, i.e., the difference in the counterfactual distributions' expected values.

The  weighted integral difference corresponding to Equation \eqref{eq:abs_varphi} is then
\begin{gather}
    L_w = \int_{\mathbb{R}} \left\{ |\varphi_1(t)|^2-|\varphi_0(t)|^2 \right\} w(t) dt,
    \label{eq:l_w}
\end{gather}
where $w(t)$ is an appropriately chosen weighting function.\footnote{Although it is clear that Equation \eqref{eq:abs_varphi} implies Equation \eqref{eq:l_w}, the converse is not generally true, so one may lose power against certain alternatives by focusing on integral differences. Alternatively, one can formulate a test based on integrals of squared differences of the modulus of characteristic functions by using a quantity such as $\mathbb{L}_w=\int_\mathbb{R}[|\varphi_1(t)|^2-|\varphi_0(t)|^2]^2w(t)dt$. In this case, since the $|\varphi_d(t)|^2$ terms in $\mathbb{L}_w$ are indeed characteristic functions, the theorems in Section \ref{sec:asym_beha} simplify and follow directly from results in \citet{meintanis2005permutation}. When $w(t)$ in $\mathbb{L}_w$ takes the form of a density of a stable distribution with parameter $\theta$, $\theta\in (0,2]$, that quantity becomes $\EX\left[\exp(-|Y(1)-Y'(1)-Y''(1)+Y'''(1)|^\theta)\right.$\linebreak$\left.-\exp(-|Y(0)-Y'(0)-Y''(0)+Y'''(0)|^\theta) \right]$ where $Y'(d)$, $Y''(d)$, and $Y'''(d)$ are independent copies of $Y(d)$, $d=0,1$. In practice, the use of this double difference, $Y(d)-Y'(d)-Y''(d)+Y'''(d)$, implies computing a large number of differences which becomes computationally expensive in the order of $O(n^4)$.
}

I summarize the previous observations in the following result.
\begin{prop}
\label{prop:l_w}
Let $Y(0)$ and $Y(1)$ be two random variables in $\mathbb{R}$. Then, for any density $w$, the quantity $L_w$ defined Equation in \eqref{eq:l_w} is equal to zero if $Y(1) \overset{d}{=} Y(0) + \tau$, for some $\tau \in \mathbb{R}$.
\end{prop}
\begin{proof}
    See Appendix \ref{app:proofs}.
\end{proof}
Proposition \ref{prop:l_w} states that whenever the null hypothesis is true, the quantity defined in Equation \eqref{eq:l_w} is zero. 

Now, let $L_\theta$ be the quantity $L_w$ where one substitutes $w(t)$ with the density of a spherical stable random variable with parameter $\theta \in (0,2]$, which I denote as $w_\theta(t)$. Following known facts about the integrals of cosine functions for this family of densities as in \citealt{chen2019some,zolotarev1981integral}, one can write Equation \eqref{eq:l_w} in the following way
\begin{gather}
    L_\theta = \EX\left[\exp(-|Y(1)-Y'(1)|^\theta)-\exp(-|Y(0)-Y'(0)|^\theta) \right].
    \label{eq:l_theta}
\end{gather}
with $Y'(0)$ and $Y'(1)$ being independent variables with the same distribution as $Y(0)$ and $Y(1)$, respectively.\footnote{Two notable cases of the spherical stable family are the standard normal and Cauchy
distributions corresponding to $\theta = 2$ and $\theta = 1$, respectively.} %
Equation \eqref{eq:l_theta} shows that the test statistic $L_\theta$ is based on the difference of the symmetrized variables $Y(0) - Y'(0)$ and $Y(1)-Y'(1)$ \citep[see][Ch.\ V]{feller2008introduction}.\footnote{To obtain $Y'(1)$, a researcher can reproduce the following steps. Given a data point within a random sample, let's say $Y_i(1)$, the researcher obtains a value for $Y'(1)$ by drawing $Y_j(1)$ from the rest of the sample with $j$ different from $i$. Since one has a random sample, $Y_i(1)$ and $Y_j(1)$ are independent and have the same distribution.}

The use of $L_\theta$ leads to the following corollary.
\begin{corollary}
    \label{co:l_theta}
  Let $Y(0)$ and $Y(1)$ be two random variables in $\mathbb{R}$. Then, for any fixed $\theta \in (0,2]$, the quantity $L_\theta$ defined in \eqref{eq:l_theta} is equal to zero if $Y(1)\overset{d}{=} Y(0) + \tau$, for some $\tau \in \mathbb{R}$. 
\end{corollary}
\begin{proof}
    See Appendix \ref{app:proofs}
\end{proof}

Both Proposition \ref{prop:l_w} and Corollary \ref{co:l_theta} show that, when the null hypothesis $H_{0}^\text{nocov}$ is true, the statistics $L_w$ and $L_\theta$ are equal to zero. However, in certain curious cases, those statistics can be zero without $H_{0}^\text{nocov}$ being true (see \citealp[p.\ 122]{lukacs1970characteristic}; \citealp[p.\ 506]{feller2008introduction}).%
Despite this fact, in the simulations of Section \ref{sec:simul}, I show that a test based on $L_\theta$ is more powerful to reject plausible alternatives than previous tests based on the characteristic function \citep{henze2005checking}. Intuitively, since a test based on $L_\theta$ does not require the estimation of $\tau$, the approximation to the large sample distribution of the recentered statistic will contain less error in the order of $O(n^{-1})$, which is the condition that the test in \citet{henze2005checking} assumes.

Finally, I introduce the finite sample version of the test quantity $L_\theta$ based on a sample of size $n$,
\begin{gather}
    \label{eq:l_n_theta}
    L_{n,\theta} = \frac{1}{n_1^2} \sum_{i:D_i=1}^n\sum_{j:D_j=1}^n  \exp(-|Y_i-Y_j|^\theta)-\frac{1}{n_0^2} \sum_{i:D_i=0}^n\sum_{j:D_j=0}^n \exp(-|Y_i-Y_j|^\theta),    \quad 0<\theta \leq 2.
\end{gather}

\subsection{Asymptotic behavior}

\label{sec:asym_beha}

In this subsection, I study the asymptotic behavior of sample analogs of the test statistic $L_w$ under both general alternatives and the null hypothesis $H_{0}^\text{nocov}$.\footnote{I offer an alternative characterization of the asymptotic distribution of the sample analog of $L_\theta$ in Appendix \ref{sec:asympt_l_theta}, which applies when the weighting scheme $w(t)$ belongs to the spherical stable family.} 

Let
\begin{gather}
    L_{n,w}  = \int_{\mathbb{R}} \left\{ |\varphi_{n,1}(t)|^2-|\varphi_{n,0}(t)|^2 \right\} w(t) dt,
    \label{eq:l_n_w}
\end{gather}
be the sample analog of $L_w$, where $\varphi_{n,1}(t)=\frac{1}{n_1}\sum_{j:D_j=1} e^{\mathrm{i}tY_{j}}$ and $\varphi_{n,0}(t)=\frac{1}{n_0}\sum_{j:D_j=0} e^{\mathrm{i}tY_{j}}$ denote the empirical characteristic functions computed on the treated and untreated samples, respectively.

A convenient space to describe the convergence in law of $L_{n,w}$ is the separable Hilbert space of measurable real-valued functions on $\mathbb{R}$ that are square integrable with respect to $w(t)$ and denoted by $\mathcal{L}_2$, with the inner product and norm in this space, respectively, defined by
\begin{gather*}
    \langle f,g \rangle_w = \int_{\mathbb{R}} f(t)g(t) w(t)dt,
    \qquad \text{and} \qquad
    \norm{f}_w = \left\{\int_\mathbb{R} [f(t)]^2w(t)dt\right\}^{1/2}.
\end{gather*}
I use similar results for $\mathcal{L}_2$-type weighted integral distances explored in, for instance, \citet{henze2005checking}, \citet{meintanis2005permutation} and \citet{chen2019some}. $L_{w,n}$ differs from the statistics in those papers slightly so; whereas $L_w$ is a difference of weighted integrals of squared characteristic functions, the statistics in those papers are weighted integrals of squared differences of characteristic functions.

I now introduce two assumptions regarding the sample size and the weighting function. Let $\xrightarrow[]{P}$ stand for convergence in probability.
\begin{assumption}[Stable treatment proportions]
\label{asm:stable_prop}
As $n\rightarrow \infty$, $n_1/n \xrightarrow{P} \pi_1 \in (0,1)$.
\end{assumption}
\begin{assumption}
    \label{asm:weight}
    Let $w(t)$ be a measurable non-negative function on $\mathbb{R}$ such that, for all $t\in \mathbb{R}$, $w(t)=w(-t)$ and $0<\int_\mathbb{R} w(t) dt<\infty$.
\end{assumption}
Assumption \ref{asm:stable_prop} specifies that any large sample has a positive fraction of control and treatment units. The symmetry of $w(t)$ around the origin stated in Assumption \ref{asm:weight} is for convenience; since one can define for any non-symmetrical weight function $w(t)$ a symmetrized version $w_1(t)= 0.5\{w(t)+w(-t)\}$ that leaves the statistic $L_{w}$ unchanged \citep[see][Equation (3), p.\ 392]{jimenez2017fast}.

In the following theorems, I describe the asymptotic distribution of $L_{n,w}$ under both general alternatives and the null hypothesis $H_{0}^\text{nocov}$. Let $\rightsquigarrow$ stands for convergence in distribution. Additionally, let $\Real(a)$ and $\Imag(a)$ be the real and imaginary parts of a complex number $a$.
\begin{theorem}
    \label{thm:l_n}
    Under Assumptions \ref{asm:random_assign}, \ref{asm:asm_randsample}, \ref{asm:stable_prop} and \ref{asm:weight}, as $n\rightarrow \infty$
    \begin{gather}
    \frac{n_0n_1}{n}\left[L_{n,w} -\int_\mathbb{R}\Delta^2_{n}(t)w(t)dt\right] \rightsquigarrow 
    \int_\mathbb{R} [\pi_0|Z_1(t)|^2 -  \pi_1|Z_0(t)|^2] w(t) dt.
    \end{gather}
where $\Delta^2_{n}(t) = \Delta^2_{n,1}(t) - \Delta^2_{n,0}(t)$, and for $d=0,1$, $\Delta^2_{n,d}(t) = \Real[\varphi_d(t)]\Real[\varphi_{n,d}(t)]+\Imag[\varphi_d(t)]\Imag[\varphi_{n,d}(t)]$ for all $t\in\mathbb{R}$. $\{Z_1(t):t\in \mathbb{R}\}$ and $\{Z_0(t):t\in \mathbb{R}\}$ are zero-mean complex Gaussian processes with covariance kernels defined, for all $s, t \in \mathbb{R}$ and $d=0,1$, by 
\begin{gather*}
    \gamma_d(s,t)  = \EX[Z_d(s)\overline{Z_d(t)}] = \varphi_d(s-t)-\varphi_d(s)\varphi_d(-t),
\end{gather*}
and complementary covariance kernel defined, for all $s,t\in \mathbb{R}$ and $d=0,1$, by
\begin{gather*}
    \gamma_d^C(s,t)  = \EX[Z_d(s)Z_d(t)] =\varphi_d(s+t)-\varphi_d(s)\varphi_d(t).
\end{gather*}
\end{theorem}
\begin{proof}
    See Appendix \ref{app:proofs}.
\end{proof}

I now introduce the following corollary that describes the limit distribution of $L_{n,w}$ under $H_{0}^\text{nocov}$. 

\begin{corollary}
    \label{thm:l_n_H0}
    Under $H_{0}^\text{nocov}$ and Assumptions \ref{asm:random_assign}, \ref{asm:asm_randsample}, \ref{asm:stable_prop} and \ref{asm:weight}, as $n\rightarrow \infty$,
    \begin{gather}
    \sqrt{\frac{n_0n_1}{n}}L_{n,w} \rightsquigarrow \int_\mathbb{R}\left[\pi_0^{1/2}\zeta_1(t)-\pi_1^{1/2} \zeta_0(t)\right]w(t)dt .
    \end{gather}
where $\{\zeta_1(t):t\in \mathbb{R}\}$ and $\{\zeta_0(t):t\in \mathbb{R}\}$ are zero-mean Gaussian processes with covariance kernels defined, for all $s, t \in \mathbb{R}$ and $d=0,1$, by 
\begin{multline*}
    \EX[\zeta_d(t)\zeta_d(s)]=\\\mathrm{Cov}\{\Real[\varphi_d(t)] \cos[tY(d)]+\Imag[\varphi_d(t)] \sin[tY(d)],\Real[\varphi_d(s)] \cos[sY(d)]+\Imag[\varphi_d(s)] \sin[sY(d)]\}.
\end{multline*}
\end{corollary}
\begin{proof}
    See Appendix \ref{app:proofs}.
\end{proof}

Corollary \ref{thm:l_n_H0} guarantees that the test based on $L_{n,w}$ is asymptotically exact in the sense that its limiting rejection probability under the null hypothesis equals the nominal level. 

The asymptotic null distribution of the statistic $L_{n,w}$ is highly non-standard. Typically, the test criteria have the same distribution as an infinite linear combination of independent chi-squared distributions and it is extremely hard to find analytically. Following the literature, I will use resampling techniques to approximate the distribution of the test statistic under the null hypothesis and general alternatives \citep{henze2005checking,meintanis2005permutation,jimenez2017fast}. 

One can extend the test based on $L_{n,w}$ to consider constant treatment effects for multiple outcomes. Specifically, one can test the hypothesis $H_0^{MO} : \mathbf{Y}(1) \overset{d}{=} \mathbf{Y}(0) + \tau $, for some $\boldsymbol{\tau} \in \mathbb{R}^p$, where $\mathbf{Y}(d)$ is a random variable in $\mathbb{R}^p$, $p$ a positive integer and $d=0,1$. The characteristic function of $\mathbf{Y}(d)$ is defined by $\varphi_d(\mathbf{t}) = \EX[\exp\{\mathrm{i}\mathbf{t}'\mathbf{Y}(d)\}]$, for $d=0,1$ and $\mathbf{t} \in \mathbb{R}^p$. Accordingly, one can redefine $L_w$ and $L_{n,w}$ to be integrals over $\mathbb{R}^p$ instead of $\mathbb{R}$ in Equations \eqref{eq:l_w} and \eqref{eq:l_n_w}.\footnote{Specifically, let $w(\mathbf{t})$ be a density on $\mathbb{R}^p$ and $\varphi_{n,d}(\mathbf{t}) = \sum_{i:D_i=d}^n\exp\{\mathrm{i}\mathbf{t}'\mathbf{Y}_i(d)\}$, $d=0,1$. One can redefine $L_w = \int_{\mathbb{R}} \left\{ |\varphi_{n,1}(\mathbf{t})|^2-|\varphi_{n,0}(\mathbf{t})|^2 \right\} w(\mathbf{t}) d\mathbf{t}$ and $L_{n,\theta}=  \frac{1}{n_1^2} \sum_{i:D_i=1}^n\sum_{j:D_j=1}^n  \exp(-||\mathbf{Y}_i-\mathbf{Y}_j||^\theta)-\frac{1}{n_0^2} \sum_{i:D_i=0}^n\sum_{j:D_j=0}^n \exp(-||\mathbf{Y}_i-\mathbf{Y}_j||^\theta),    \ 0<\theta \leq 2$, with $||\cdot||$ the Euclidean norm.}


\subsection{The covariates case}
\label{sec:cov}
A second hypothesis of interest is whether there is remaining variation in treatment effects after removing the variation predicted by predetermined characteristics as shown in Equation \eqref{eq:h0}. Testing for this hypothesis is equivalent to testing for the equality, across treatment groups, of the distributions of the conditional expectation errors:
\begin{gather}
   H_0: \varepsilon(1) \overset{d}{=} \varepsilon(0),
\end{gather}
where $\varepsilon(1) \equiv Y(1)-\EX[Y(1)|\mathbf{X}]$ and $\varepsilon(0)\equiv Y(0)-\EX[Y(0)|\mathbf{X}]$ are the conditional expectation errors of the treated and control groups, and $\EX[Y(d)|\mathbf{X}]$ is the conditional expectation of $Y(d)$ given $\mathbf{X}$, $d=0,1$. By recentering the potential outcomes using the conditional expectations, one would remove, in a mean-squared sense, the variation in the treatment effects accounted for predetermined characteristics.\footnote{\citet[see p.\ 662-3]{ding2016randomization} used a similar procedure in a finite population setting where, instead of using the conditional expectation, they suggest using a linear regression projection $\mathbf{X}'\boldsymbol{\beta}$.}

The following regularity condition guarantees that the conditional expectation function and the best linear predictor of $Y(d)$ given $\mathbf{X}$ exists.
\begin{assumption} 
    \label{asm:cef}
    \leavevmode
    For $d=0,1$, $E[|Y(d)|]<\infty$, and, $\EX[\mathbf{X}Y(d)]$ and $\EX[\mathbf{X}\mathbf{X}']$ exist and have finite elements.
\end{assumption}
In Assumption \ref{asm:cef}, $E[|Y(d)|]<\infty$ guarantees that the conditional function $\EX[Y|X]$ exists, whereas the existence and finiteness of the elements of  $\EX[\mathbf{X}Y(d)]$ and $\EX[\mathbf{X}\mathbf{X}']$ guarantees that the best linear predictor of $Y(d)$ given $\mathbf{X}$ exist.%
\footnote{Note that the second condition in Assumption \ref{asm:cef} is weaker than assuming that $\EX[\mathbf{X}\mathbf{X}']$ is invertible, which guarantees that the linear projection coefficient is unique \citep[see][p.\ 36]{hansen2021econometrics}.}
The best linear predictor of $Y(d)$ given $\mathbf{X}$ is denoted by
\begin{gather}
    \mathcal{P}(Y(d)|\mathbf{X}) = \mathbf{X}'\beta_d, \qquad d=0,1,
\end{gather}
where $\beta_d = \EX[\mathbf{X}\mathbf{X}']^{-}\EX[\mathbf{X}Y(d)]$ is the linear projection coefficient, and $A^{-}$ is the generalized inverse of matrix $A$. 

We can use the conditional expectation errors, $\varepsilon(1)$ and $\varepsilon(0)$, to compute the following $\mathcal{L}_2$-type distance based on the characteristic function
\begin{gather}
    \mathcal{D}_{w} = \int_\mathbb{R} [\varphi_{\varepsilon(1)}(t)-\varphi_{\varepsilon(0)}(t)]^2 w(t)dt,
    \label{eq:d_w}
\end{gather}
where $\varphi_{\varepsilon(d)}(t)=\EX[\exp\{\mathrm{i}t\varepsilon(d)]$ is the characteristic function of the error $\varepsilon(d)$, $d=0,1$. The distance in Equation \eqref{eq:d_w} has been previously used in \citet{meintanis2005permutation}, \citet{chen2019some} and \citet{rivas2019two}.

I formulate the following proposition that resembles Proposition \ref{prop:l_w} for instances when the researcher has available covariates.
\begin{prop}
\label{prop:d_w}
Let $Y(0)$ and $Y(1)$ be two random variables in $\mathbb{R}$. Under Assumptions \ref{asm:weight} and \ref{asm:cef},  the quantity $\mathcal{D}_w$ defined in Equation \eqref{eq:d_w} is equal to zero if and only if $H_0$ is true.
\end{prop}
\begin{proof}
    See Appendix \ref{app:proofs}.
\end{proof}


One possibility to estimate the functions $\varphi_{\varepsilon(0)}(t)$ and $\varphi_{\varepsilon(1)}(t)$ is to calculate the residuals from a nonparametric regression
\begin{gather}
    \widehat{\varepsilon}_i(d)=Y_i-\widehat{m}_d(X_i),    \quad d=0,1,
\end{gather}
where $\widehat{m}_d(x)$ is a nonparametric estimator of ${m}_d(x)\equiv\EX[Y(d)|\mathbf{X}=x]$ \citep[see][]{chen2007large,chen2008semiparametric,crump2008nonparametric}. In particular, I use the following kernel estimators for the conditional expectation function ${m}_d(x)$
\begin{gather*}
    \widehat{m}_d(x) = \sum_{i:D_i=d}^{n} \omega_{in}^{(d)}(x)Y_i, \quad x \in S,
    \intertext{with associated weights $\omega_{in}^{(d)}(x)$ given by}
    \omega_{in}^{(d)}(x) = \frac{K_{h_d}(\mathbf{X}_i-\mathbf{x})}{\sum_{s:D_s=D_i=d}^{n}K_{h_d}(\mathbf{X}_s-\mathbf{x})}, \quad \mathbf{x}\in S,
\end{gather*}
and $K(v)$ is a kernel and $h$ is the bandwidth such that $K_{h}(v)=\frac{1}{h}K(v/h)$. For simplicity, I assumed that the same kernel function, $K$, is used for both treatment and control groups. 

The following regularity conditions ensure that the Nadaraya-Watson type kernel estimator $\widehat{m}_d$ will be a consistent estimator for the conditional expectation functions $m_d(x)$ \citep{hansen2008uniform}.
\begin{assumption}[Conditions for uniformly consistent estimation of $m_d(x)$]
    \label{asm:kernel} For $d=0,1$,
    \begin{enumerate}[(a)]
    
    \setlength{\itemsep}{0pt}
    \setlength{\parskip}{0pt}
    \setlength{\parsep}{0pt} 
    
    \item $|K(u)|\leq \overline{K}<\infty$ and $\int_{\mathbb{R}^q}|K(u)|du\leq M < \infty$.
    \item $\EX[Y(d)^s]<\infty$ for some $s>2$.
    \item $\mathbf{X}$ has density $f$ such that $\sup_{\mathbf{x}} f(\mathbf{x}) \leq B_0<\infty$ and $\sup_{\mathbf{x}} \EX[|Y(d)|^s|\mathbf{X}=\mathbf{x}]f(\mathbf{x}) \leq B_1<\infty$.
    \item The second derivatives of $f(\mathbf{x})$ and $f(\mathbf{x})m_d(\mathbf{x})$ are uniformly continuous and bounded.
    \item $K(u):\mathbb{R}^d\rightarrow\mathbb{R}$ is a symmetric kernel such as
    \begin{itemize}
        \item For some $\Lambda_1<\infty$ and $L<\infty$, either $K(u)=0$ for $||u||>L$ and for all $u,u' \in \mathbb{R}^q$, $|K(u)-K(u')|\leq \Lambda_1 ||u-u'||$ or 
        \item $K(u)$ is differentiable with $|\partial K(u)/\partial u| \leq \Lambda_1$, and for some $v>1$, $|\partial K(u)/\partial u| \leq \Lambda_1 ||u||^{-v}$ for $||u||>L$. 
    \end{itemize}
    \item $h \rightarrow 0 $ and $(\ln n/nh^q)^{1/2}\rightarrow 0$ as $n\rightarrow 0$
    \end{enumerate}
\end{assumption}


After estimating the residuals, I can compute the sample counterpart of $\mathcal{D}_w$ as
\begin{gather}
    \mathcal{D}_{n,w} = \int_\mathbb{R} [\varphi_{\widehat{\varepsilon}(1),n}(t)-\varphi_{\widehat{\varepsilon}(0),n}(t)]^2 w(t)dt
\end{gather}
where $\varphi_{\varepsilon(1),n}(t)= \frac{1}{n_1} \sum_{j=1}^n D_j \exp\{\mathrm{i}t\widehat{\varepsilon}_j\}$ and $\varphi_{\varepsilon(0),n}(t)= \frac{1}{n_0} \sum_{j=1}^n (1-D_j) \exp\{\mathrm{i}t\widehat{\varepsilon}_j\}$ are the empirical characteristic functions of the residuals $\widehat{\varepsilon}(1)$ and $\widehat{\varepsilon}(0)$, respectively. 
Large values of $\mathcal{D}_{n,w}$ indicate that there is treatment effect variation after accounting for covariates. Similar to the case without covariates, I can compute critical values for the test associated with $\mathcal{D}_{n,w}$ using resampling methods, which I describe in Section \ref{sec:simul}. 

The following two theorems are the basis for the asymptotic validity and consistency of the test based on $\mathcal{D}_{n,w}$.

\begin{theorem}[Stochastic limit of $\mathcal{D}_{n,w}$]
    \label{thm:d_w_stoch_lim}
    Under Assumptions \ref{asm:stable_prop}--\ref{asm:kernel}, as $n\rightarrow\infty$,
    \[\mathcal{D}_{n,w}\xrightarrow[]{P}  \mathcal{D}_w =\norm{\varphi_{\varepsilon(1)}(t)-\varphi_{\varepsilon(0)}(t)}_w^2.\]
\end{theorem}
\begin{proof}
    See Appendix \ref{app:proofs}.
\end{proof}

Theorem \ref{thm:d_w_stoch_lim} shows that the sample analog $\mathcal{D}_{n,w}$ is consistent for the population quantity $\mathcal{D}_{w}$. Theorem \ref{thm:d_w_stoch_lim} together with Proposition \ref{prop:d_w} make it possible to use $\mathcal{D}_{n,w}$ as the basis for a consistent test based on the empirical characteristic function. In view of Proposition \ref{prop:d_w}, the quantity $\mathcal{D}_w$ is equal to zero if and only if the null hypothesis $H_0$ holds true provided that the weight function $w$ is positive with probability 1. Consequently, large values of $\mathcal{D}_{n,w}$ should imply a rejection of the null hypothesis $H_0$. 

The following theorem shows that a standardized $\mathcal{D}_{n,w}$ has a limiting distribution as the sample size increases. 
\begin{theorem}
    \label{thm:d_w_asym}
    Under $H_0$ and Assumptions \ref{asm:stable_prop}--\ref{asm:kernel}, as $n\rightarrow\infty$
    \begin{gather}
        \frac{n_0n_1}{n}\mathcal{D}_{n,w} \rightsquigarrow \norm{Z^{\text{cv}}}_w^2
    \end{gather}
where $\{Z^{\text{cv}}(t),t\in\mathbb{R}\}$ is a zero-mean Gaussian process on $\mathcal{L}_2$ with covariance kernel $	\varrho_{0}(s,t) = \text{Cov}_0\{Z_0(\varepsilon;s)Z_0(\varepsilon;t)\}$ and
$Z_0(\varepsilon;t) = \cos(t\varepsilon)+\sin(t\varepsilon)+t\varepsilon(\Imag[\varphi_\varepsilon(t)]-\Real[\varphi_\varepsilon(t)])-\Real[\varphi_\varepsilon(t)]-\Imag[\varphi_\varepsilon(t)].$
\end{theorem}
\begin{proof}
    See Appendix \ref{app:proofs}.
\end{proof}

As with the limiting distribution of statistic $L_{n,w}$, the asymptotic distribution of $\mathcal{D}_{n,w}$ is highly non-standard \citep[see Remark 1 in][]{rivas2019two}. Following the literature, I will show the use of resampling techniques to approximate the null distribution of this test statistic in the next section.

When the weighting scheme belongs to the spherical stable density family with parameter $\theta$, it is straightforward to define the corresponding weighted integrals $\mathcal{D}_\theta$ and $\mathcal{D}_{n,\theta}$ as in the case of $L_\theta$ based on Equations \eqref{eq:l_theta} and \eqref{eq:l_n_w}. Similar results to those in Corollary \ref{co:l_theta} and Theorems \ref{thm:d_w_stoch_lim} and \ref{thm:d_w_asym} based on the quantity $\mathcal{D}_\theta$ and the statistic $\mathcal{D}_{n,\theta}$ follow. For the sake of brevity, I omit those results.

Finally, I introduce the finite sample version of the test quantity $\mathcal{D}_\theta$ based on a sample of size $n$,
\begin{align}
    \notag
    \mathcal{D}_{n,2} = \frac{1}{n_0^2}& \sum_{i=1}^{n}\sum_{j=1}^{n}(1-D_i)(1-D_j) \exp(-|\widehat{\varepsilon}_i-\widehat{\varepsilon}_j|^\theta)+\frac{1}{n_1^2} \sum_{i=1}^{n} \sum_{j=1}^{n}  D_iD_j\exp(-|\widehat{\varepsilon}_i-\widehat{\varepsilon}_j|^\theta)
    \notag
    \\
    &-\frac{2}{n_0n_1} \sum_{i=1}^{n}\sum_{j=1}^{n} (1-D_i)D_j\exp(-|\widehat{\varepsilon}_i-\widehat{\varepsilon}_j|^\theta).
    \label{eq:d_n_theta}
\end{align}


\section{Simulations}
\label{sec:simul}
I now describe a simulation study that confirms the validity of the tests based on $L_{n,w}$ [see Equation \eqref{eq:l_n_w}]
and assesses their power under a range of plausible scenarios. To compare these results with previous studies, I follow the data generating processes suggested in \citet{ding2016randomization}, \citet{koenker2002inference}, and \citet{chernozhukov2005subsampling}.%
\footnote{\citet{ding2016randomization} kindly provide simulation code in their paper's supplemental material.} 

When implementing the test based on $L_{n,\theta}$ in Equation \eqref{eq:l_n_theta}, I select the density function of the spherical stable distribution with parameter $\theta=2$, which corresponds to the normal density, and rewrite $L_{n,\theta}$ in the following way:
\begin{align}
    L_{n,2} = \frac{1}{n_1^2} \sum_{i=1}^{n} \sum_{j=1}^{n}  D_iD_j\exp(-|Y_i-Y_j|^2)-\frac{1}{n_0^2}& \sum_{i=1}^{n}\sum_{j=1}^{n}(1-D_i)(1-D_j) \exp(-|Y_i-Y_j|^2).
    \label{eq:l_mine}
\end{align}

To compute critical values for the test based on $L_{n}$, I use the bootstrap, which does not require the estimation of the constant treatment effect, and a permutation test, which does require that estimation. Intuitively, the bootstrap and permutation tests provide good approximations to the asymptotic distribution of $L_{n,\theta}$ since the empirical characteristic function is a sample average of a smooth function of the empirical process. 

Besides $L_{n,2}$, I include simulation results for the test statistic in \citet{henze2005checking}:
\begin{align}
    L_{n}^{HKZ} = \frac{1}{n_0^2}& \sum_{i=1}^{n}\sum_{j=1}^{n}(1-D_i)(1-D_j) \exp(-|Y_i-Y_j|^2)+\frac{1}{n_1^2} \sum_{i=1}^{n} \sum_{j=1}^{n}  D_iD_j\exp(-|Y_i-Y_j|^2)
    \notag
    \\
    &-\frac{2}{n_0n_1} \sum_{i=1}^{n}\sum_{j=1}^{n} (1-D_i)D_j\exp(-|Y_i+\widehat{\tau}-Y_j|^2),
    \label{eq:l_khz}
\end{align}
where $\widehat{\tau}=\frac{1}{n_1}\sum_{i:D_i=1}Y_{i}-\frac{1}{n_0}\sum_{i:D_i=0}Y_{i}$ is the classic differences-in-means estimator of $\tau$. The test based on $L_{n}^{HKZ}$ requires that $\widehat{\tau}-\tau$ has a representation as the difference-in-means of an appropriate measurable function with mean zero and finite variance, which boils down to $\sqrt{n}(\widehat{\tau}-\tau)=O_P(1)$ when $E[Y(0)^2]<\infty$ and $E[Y(1)^2]<\infty$ \citep{henze2005checking}. To compute critical values for the test based on $L_{n}^{HKZ}$, I follow \citet{henze2005checking} and use the bootstrap and a permutation test.

The most striking difference between the two test statistics $L_{n,2}$ and $L_{n}^{HKZ}$ is that $L_{n}^{HKZ}$ depends on the estimation of the average treatment effect, whereas $L_{n,2}$ does not. Furthermore, as suggested by Equation \eqref{eq:l_n_theta} and to possibly increase power, one can generalize $L_{n,2}$ to use other parameters of the spherical stable distribution as long as $\theta \in (0,2]$. Although, the statistic $L_{n}^{HKZ}$ can also be generalized in a similar way, the formulation in \citet{henze2005checking} does not easily introduces this possibility.

\subsection{Validity results}

First, I examine the various methods under the null hypothesis of a constant treatment effect (no covariates) as specified in the following model:
\begin{equation}
\begin{aligned}
    Y_i(0) &= \varepsilon_i,
    \\
    Y_i(1) &= Y_i(0) + \tau,
\end{aligned}
\label{eq:sim_h0}
\end{equation}
with $i = 1, \dots, n$. I assess the methods based on $L_{n,2}$ and $L_{n}^{HKZ}$ for the following distributions:  $\varepsilon_i$ distributed standard normal, $t_5$, standard exponential and log-normal; each with a constant treatment effect of 1 unit ($\tau = 1$) as suggested in \citet{koenker2002inference}, \citep{chernozhukov2005subsampling}, and \citet{ding2016randomization}.

To assess validity, I replicate the following procedure 5,000 times for each combination of a test statistic, a sample size and an underlying distribution:

\begin{enumerate}
    \item generate a random sample $[Y_1(0),Y_1(1)],\dots,[Y_{n}(0),Y_{n}(1)]$ from the underlying distribution, assuming a constant treatment effect;
    \item \label{item_2} randomly assign treatment $D_1,\dots,D_n$ and obtain observed outcomes $Y_1,\dots,Y_n$;
    \item calculate the test statistics $L_{n,2}$ and $L_{n}^{HKZ}$;
    \item calculate a critical value and reject $H_0^\text{nocov}$ using either of the following methods:
    \begin{enumerate}
    \item A bootstrap test for ${L}_{n,2}$, 
    \begin{enumerate}
        \item draw 2,000 samples with replacement from the sample $(D_1,Y_1),\dots,(D_n,Y_n)$;
        \item calculate the corresponding 2,000 realizations $L_{n,2}^{B}(j)$, $1\leq j \leq 2,000$, of the bootstrap statistic $L_{n,2}^{B}$;
        \item recenter each realization of the bootstrap statistic $L_{n,2}^{B}$ using the calculated value $L_{n,2}$, $L_{n,2}^{B}(j)-L_{n,2}$, $1\leq j \leq 2,000$ (as Theorem \ref{thm:l_n} suggests);
        \item reject $H_{0}^\text{nocov}$ if $L_{n,2}$ computed on $(D_1,Y_1),\dots,(D_n,Y_n)$ is either lower than the empirical $2.5\%$-quantile of $L_{n,2}^{B}(j)-L_{n,2}$ or greater than the empirical $97.5\%$-quantile of $L_{n,2}^{B}(j)-L_{n,2}$.
    \end{enumerate}
    \item Bootstrap and permutation tests for $L_{n}^{HKZ}$ (as in \citealp{henze2005checking}), 
    \label{it:perm_henze}
    \begin{enumerate}
        \item compute $Y_i'= (1-D_i)(Y_i+\widehat{\tau})+D_iY_i$.
        \item draw 2,000 samples with (without) replacement from the sample $Y_1',\dots,Y_n'$, and draw 2,000 corresponding treatment vectors of the form $D_1',\dots,D_n'$ according to the treatment assignment rule in \ref{item_2};
        \item calculate the corresponding 2,000 realizations $L_{n}^{HKZ,B}(j)$ [$L_{n}^{HKZ,P}(j)$], $1\leq j \leq 2,000$, of the bootstrap (permutation) statistic $L_{n}^{HKZ,B}$ ($L_{n}^{HKZ,P}$);
        \item reject $H_{0}^\text{nocov}$ if $L_{n}^{HKZ}$ computed on $(D_1,Y_1),\dots,(D_n,Y_n)$ exceeds the empirical $95\%$-quantile of $L_{n}^{HKZ,B}(j)$ [$L_{n}^{HKZ,P}(j)$].
    \end{enumerate}
    \item A permutation test for $L_{n,2}$ using the steps in \ref{it:perm_henze} (sampling without replacement).\footnote{It can be shown that such permutation test for the statistic $L_{n,w}$ is a randomization inference test (see Proposition \ref{prop:perm} in Appendix \ref{app:proofs}).}
    \end{enumerate}
\end{enumerate}

As suggested in \citet{ding2016randomization}, there is a nuisance parameter problem when implementing permutation tests based on the Kolmogorov-Smirnoff statistic in the form of the average treatment effect, $\tau$. The bootstrapped test based on $L_{n,2}$ does not directly depend on this parameter, so it does not require the estimation of this parameter, whereas all implementations of the test in \citet{henze2005checking} does require the estimation of such parameter. On the other hand, the permutation version of the test based on $L_{n,2}$ does require the estimation of the average treatment effect to calculate an approximate science table, i.e., the table that lists all of the pairs of potential outcomes for every unit. 

Table \ref{tb:size} shows the rejection rates of 5\%-level tests for the tests based on the characteristic function, the plug-in and confidence-interval Fisher randomization tests \citep[][]{ding2016randomization}, the martingale transformation test \citep{CHUNG2021148} and the subsampling quantile regression test \citep{chernozhukov2005subsampling}, for each sample size and data-generating process. The actual level of the bootstrap and permutation procedures based on ${L}_{n,2}$ is close to the nominal level of 5\% for samples sizes as small as 200 observations. The bootstrap and permutation methods based on $L_{n}^{HKZ}$ exhibit severe overrejection when the underlying distribution is log-normal. However, this overrejection is not as severe as the one exhibited by the plug-in Fisher randomization test. The confidence-interval Fisher randomization, the subsampling and the martingale transformation tests seem to control size in samples as small as 50 observations.

To overcome the overrejection of the characteristic-function based tests in samples with less than 200 observations, I implement the confidence interval correction in \citet{berger1994p} and \citet{ding2016randomization}. Specifically, I repeat the steps in the permutation test (see item \ref{it:perm_henze} in the simulation procedure above) for $m$ values of $\tau$ chosen from a valid $99.9\%$-level confidence interval for $\tau$. In practice, this method may require performing permutation tests for a large $m$. When applying this test to data generated under constant treatment effects, as in Equation \eqref{eq:sim_h0}, with $m=21$, the confidence interval permutation test based on the characteristic function (either $L_{n,2}$ or $L_{n}^{HKZ}$) controls the probability of rejecting a true null in samples as small as 50 observations. 

\begin{table}
\caption{Size of 5\%-level tests under $H_{0}^\text{nocov}$ (in percentage points)}
\centering
\begin{ThreePartTable}
\label{tb:size}
%
\begin{tabular}[t]{rrrrrrrrrrr}
\toprule
\multicolumn{1}{c}{ } & \multicolumn{3}{c}{$L_{n,2}$} & \multicolumn{3}{c}{$L_n^{HKZ}$} & \multicolumn{2}{c}{FRT} & \multicolumn{1}{c}{MT} & \multicolumn{1}{c}{Subs} \\
\cmidrule(l{3pt}r{3pt}){2-4} \cmidrule(l{3pt}r{3pt}){5-7} \cmidrule(l{3pt}r{3pt}){8-9} \cmidrule(l{3pt}r{3pt}){10-10} \cmidrule(l{3pt}r{3pt}){11-11}
\multicolumn{1}{c}{$n$} & \multicolumn{1}{c}{Boot} & \multicolumn{1}{c}{Perm} & \multicolumn{1}{c}{CI} & \multicolumn{1}{c}{Boot} & \multicolumn{1}{c}{Perm} & \multicolumn{1}{c}{CI} & \multicolumn{1}{c}{CI} & \multicolumn{1}{c}{PI} & \multicolumn{2}{c}{ } \\
\midrule
\addlinespace[0.3em]
\multicolumn{11}{l}{\textbf{Panel A: Exponential outcomes}}\\
\hspace{1em}50 & 10.4 & 7.1 & 4.3 & 7.2 & 8.6 & 3.3 & 2.2 & 12.9 & 1.8 & 1.1\\
\hspace{1em}80 & 9.0 & 6.9 & 3.5 & 7.5 & 8.7 & 3.5 & 3.1 & 12.0 & 1.6 & 1.3\\
\hspace{1em}100 & 6.4 & 4.3 & 2.0 & 5.0 & 5.9 & 2.7 & 1.8 & 9.6 & 1.6 & 1.7\\
\hspace{1em}200 & 5.5 & 5.2 & 3.0 & 5.4 & 5.6 & 3.8 & 3.1 & 9.2 & 1.6 & 1.7\\
\hspace{1em}1000 & 4.7 & 4.5 & 2.3 & 4.5 & 5.0 & 3.2 & 4.3 & 6.6 & 0.9 & 2.3\\
\addlinespace[0.3em]
\multicolumn{11}{l}{\textbf{Panel B: Log-normal outcomes}}\\
\hspace{1em}50 & 8.2 & 7.6 & 3.6 & 11.7 & 15.7 & 3.4 & 2.0 & 19.6 & 2.2 & 1.5\\
\hspace{1em}80 & 8.2 & 7.5 & 3.3 & 10.7 & 13.3 & 3.6 & 3.4 & 16.9 & 2.8 & 0.5\\
\hspace{1em}100 & 7.6 & 7.7 & 4.0 & 9.0 & 10.3 & 3.5 & 2.9 & 15.6 & 2.5 & 1.0\\
\hspace{1em}200 & 5.0 & 5.1 & 3.6 & 7.0 & 8.1 & 4.2 & 3.8 & 10.2 & 3.2 & 0.5\\
\hspace{1em}1000 & 5.3 & 5.7 & 3.6 & 5.4 & 6.1 & 3.8 & 4.8 & 6.8 & 2.7 & 0.6\\
\addlinespace[0.3em]
\multicolumn{11}{l}{\textbf{Panel C: Normal  outcomes}}\\
\hspace{1em}50 & 5.3 & 4.2 & 2.2 & 4.4 & 5.2 & 1.4 & 0.7 & 4.6 & 3.7 & 1.2\\
\hspace{1em}80 & 6.1 & 5.3 & 2.9 & 4.7 & 5.8 & 2.4 & 1.4 & 4.2 & 3.6 & 3.0\\
\hspace{1em}100 & 5.1 & 4.8 & 2.8 & 5.3 & 5.6 & 2.9 & 1.3 & 4.3 & 3.2 & 1.8\\
\hspace{1em}200 & 5.5 & 5.0 & 3.3 & 5.7 & 5.4 & 3.5 & 2.9 & 5.6 & 3.9 & 4.1\\
\hspace{1em}1000 & 5.0 & 6.0 & 2.1 & 6.1 & 5.4 & 3.6 & 3.7 & 5.0 & 3.2 & 5.1\\
\addlinespace[0.3em]
\multicolumn{11}{l}{\textbf{Panel D: $t_5$  outcomes}}\\
\hspace{1em}50 & 6.2 & 5.0 & 3.2 & 3.7 & 6.4 & 1.0 & 0.8 & 5.1 & 3.0 & 0.7\\
\hspace{1em}80 & 5.9 & 5.1 & 2.9 & 4.2 & 6.3 & 2.8 & 1.6 & 5.1 & 4.1 & 0.6\\
\hspace{1em}100 & 6.5 & 5.5 & 3.0 & 4.3 & 5.4 & 2.4 & 1.4 & 4.4 & 3.5 & 2.2\\
\hspace{1em}200 & 4.9 & 4.4 & 2.7 & 3.7 & 4.4 & 2.9 & 2.1 & 4.7 & 2.6 & 1.7\\
\hspace{1em}1000 & 4.5 & 4.0 & 2.4 & 4.4 & 5.1 & 3.3 & 2.6 & 4.3 & 2.4 & 1.9\\
\bottomrule
\end{tabular}
\begin{tablenotes}
\item{\small Estimates are based on 5,000 replications, which imply a simulation standard error of approximately 0.3 percentage points. Replicated potential outcomes follow a constant treatment effect model, as in \citet{ding2016randomization}, and specified by the equations $Y_i(0)=\varepsilon_i$, $Y_i(1)=1+Y_i(0)$ with $\varepsilon_i$ distributed standard exponential, log-normal, standard normal and $t_5$; half of the units were randomly assigned to treatment. $L_{n,2}$ and $L_{n}^{KHZ}$ correspond to the tests based on the statistics in Equations \eqref{eq:l_mine} and \eqref{eq:l_khz}, respectively. CI corresponds to the confidence interval permutation test that performs a permutation test for 21($=m$) values of $\tau$ in a 99.9\%-level confidence interval for $\tau$ \citep{berger1994p, ding2016randomization}.
FRT CI and and FRT PI stands for the confidence-interval and plug-in Fisher randomization test in \citet{ding2016randomization}; MT stands for the martingale transformation test in \citet{CHUNG2021148}; and subsampling stands for the subsampling quantile regression test in \cite{chernozhukov2005subsampling}. Boot stands for bootstrapping, and Perm stands for permutation.
}
\end{tablenotes}
\end{ThreePartTable}
\end{table}

\subsection{Power simulations}
To assess the power of these methods under selected alternatives, I adopt the simulation study in \citet{koenker2002inference}, \citet{chernozhukov2005subsampling} and \citet{ding2016randomization}. For these simulations, I repeatedly generate data with different levels of treatment effect heterogeneity, denoted by $\sigma_\tau$, and I estimate the probability that a method would reject the null hypothesis of constant treatment effect at 5\% nominal level given draws of data and random treatment assignment,
\begin{equation}
    \begin{aligned}
   & Y_i(0)  =\varepsilon_i,
    \\
&    \tau_i  = 1 + \sigma_\tau Y_i(0),
    \\
 &   Y_i(1)  = Y_i(0) + \tau_i,
    \end{aligned}
\label{eq:sim_h1}
\end{equation}
for each $i=1,\dots,n$, with $\varepsilon_i$ distributed either standard normal, $t_5$, standard exponential or log-normal, each with $\sigma_\tau=0.2$ or $\sigma_\tau=0.5$.\footnote{If $\varepsilon_i \sim \mathcal{N}(0,1)$, the $F$-test is the uniformly most powerful test \citep{cox1984interaction}.}

Table \ref{tb:power} shows the rejection rates of $H_{0}^\text{nocov}$ for those tests that control simulated size in Table \ref{tb:size}, specifically, the confidence-interval permutation versions of the tests based in $L_{n,2}$ and $L_{n}^{HKZ}$, the confidence-interval Fisher randomization test, the martingale transformed test, and the subsampling quantile regression test. Particularly, the simulations show that the confidence interval permutation test based on $L_{n,2}$ has higher power than the competing tests in log-normal, normal and $t_5$ distributions. In exponential outcomes, the confidence-interval test based on $L_{n,2}$ tends to be more powerful in samples with 100 or less observations and with high heterogeneity ($\sigma_\tau=0.5$). Surprisingly, the martingale transformation test seems to be overly conservative. 

To compare power across tests with different sizes, I compute size-adjusted
power using the simulated percentiles of the corresponding test statistics for each underlying distribution and sample size. Table \ref{tab:power_adjusted} shows the size-adjusted power of the tests based on statistics $L_{n,2}$ and $L_{HKZ}$, the quantile regression test \citep{chernozhukov2005subsampling}, the martingale transformed test \citep{CHUNG2021148}, and the FRTCI statistic (Kolmogorov-Smirnoff statistic). These calculations presume the availability of the critical values that correspond to each of the underlying distributions and sample sizes. The test based on $L_{n,2}$ has higher power than the competing tests when the distributions are symmetric, i.e., the standard normal and the $t_5$ distributions. In log-normal outcomes, the martingale-transformed statistic has higher power than the competing tests. The tests based on $L_{n,2}$ and the FRTCI tend to have higher power for exponential outcomes. Finally, within the statistics based on the characteristic function, the test based on $L_{n,2}$ appears to have higher power than the tests based on $L_{n}^{HKZ}$.


\begin{table}

\caption{\label{tb:power}Power of 5\%-level tests when $H_0^\text{nocov}$
      is false (in percentage points)}
\centering
        \begin{ThreePartTable}

\begin{tabular}[t]{rrrrrrrrrrr}
\toprule
\multicolumn{1}{c}{$n$} & \multicolumn{5}{c}{$\sigma_\tau=0.2$} & \multicolumn{5}{c}{$\sigma_\tau=0.5$} \\
\cmidrule(l{3pt}r{3pt}){1-1} \cmidrule(l{3pt}r{3pt}){2-6} \cmidrule(l{3pt}r{3pt}){7-11}
\multicolumn{1}{c}{ } & \multicolumn{1}{c}{$L_{n,2}$} & \multicolumn{1}{c}{$L_n^{HKZ}$} & \multicolumn{1}{c}{FRTCI} & \multicolumn{1}{c}{MT} & \multicolumn{1}{c}{Subs} & \multicolumn{1}{c}{$L_{n,2}$} & \multicolumn{1}{c}{$L_n^{HKZ}$} & \multicolumn{1}{c}{FRTCI} & \multicolumn{1}{c}{MT} & \multicolumn{1}{c}{Subs}\\
\midrule
\addlinespace[0.3em]
\multicolumn{11}{l}{\textbf{Panel A: Exponential outcomes}}\\
\hspace{1em}50 & 5.0 & 5.1 & 3.0 & 1.0 & 1.6 & 15.3 & 13.9 & 8.9 & 0.7 & 6.2\\
\hspace{1em}100 & 6.9 & 8.1 & 6.8 & 1.9 & 3.5 & 33.5 & 30.3 & 30.5 & 0.6 & 17.5\\
\hspace{1em}400 & 30.3 & 29.2 & 35.9 & 0.4 & 16.4 & 91.5 & 91.1 & 95.2 & 0.1 & 83.6\\
\hspace{1em}800 & 54.3 & 53.3 & 66.5 & 0.0 & 39.6 & 99.8 & 99.8 & 99.9 & 0.0 & 99.4\\
\hspace{1em}1000 & 63.9 & 62.6 & 76.6 & 0.1 & 48.4 & 99.9 & 100.0 & 100.0 & 0.0 & 99.9\\
\addlinespace[0.3em]
\multicolumn{11}{l}{\textbf{Panel B: Log-normal outcomes}}\\
\hspace{1em}50 & 5.3 & 4.1 & 2.1 & 2.1 & 1.7 & 12.2 & 8.1 & 3.6 & 1.3 & 2.9\\
\hspace{1em}100 & 8.1 & 6.8 & 5.7 & 2.1 & 1.7 & 26.5 & 21.2 & 17.6 & 1.5 & 5.7\\
\hspace{1em}400 & 22.5 & 16.5 & 17.1 & 2.7 & 3.7 & 82.5 & 67.1 & 67.5 & 0.1 & 31.8\\
\hspace{1em}800 & 37.6 & 28.2 & 33.9 & 2.0 & 8.8 & 98.5 & 91.8 & 93.2 & 0.0 & 71.9\\
\hspace{1em}1000 & 50.2 & 38.3 & 42.2 & 0.7 & 12.0 & 99.6 & 97.0 & 98.2 & 0.0 & 84.2\\
\addlinespace[0.3em]
\multicolumn{11}{l}{\textbf{Panel C: Normal  outcomes}}\\
\hspace{1em}50 & 5.0 & 3.7 & 0.7 & 4.6 & 2.5 & 25.6 & 14.1 & 3.6 & 7.8 & 10.7\\
\hspace{1em}100 & 13.9 & 7.0 & 3.2 & 6.6 & 7.0 & 60.9 & 40.3 & 18.5 & 15.2 & 39.4\\
\hspace{1em}400 & 58.4 & 40.1 & 22.9 & 18.9 & 42.5 & 99.9 & 98.5 & 92.7 & 60.8 & 97.9\\
\hspace{1em}800 & 87.7 & 71.3 & 47.5 & 34.9 & 70.5 & 100.0 & 100.0 & 100.0 & 91.2 & 100.0\\
\hspace{1em}1000 & 94.1 & 82.1 & 61.9 & 43.3 & 82.4 & 100.0 & 100.0 & 100.0 & 96.7 & 100.0\\
\addlinespace[0.3em]
\multicolumn{11}{l}{\textbf{Panel D: $t_5$  outcomes}}\\
\hspace{1em}50 & 5.4 & 2.9 & 0.7 & 3.6 & 1.6 & 22.6 & 8.8 & 2.3 & 6.3 & 6.5\\
\hspace{1em}100 & 11.3 & 6.5 & 2.5 & 6.1 & 4.8 & 48.8 & 25.7 & 10.6 & 14.2 & 21.5\\
\hspace{1em}400 & 43.7 & 24.2 & 13.4 & 14.9 & 14.6 & 98.7 & 90.7 & 70.7 & 52.3 & 78.0\\
\hspace{1em}800 & 73.5 & 49.8 & 30.3 & 26.9 & 32.3 & 100.0 & 100.0 & 97.5 & 81.4 & 97.8\\
\hspace{1em}1000 & 83.6 & 61.9 & 38.1 & 35.8 & 40.0 & 100.0 & 100.0 & 99.5 & 92.1 & 99.2\\
\bottomrule
\end{tabular}
    \begin{tablenotes}
        \item{\small Estimates are based on 5,000 replications, which imply a simulation standard error of approximately 0.3 percentage points. The table shows the simulated power for the confidence-interval characteristic-function test in this paper's $L_{n,2}$ (Equation \eqref{eq:l_mine}), and \citet{henze2005checking}'s $L_n^{HKZ}$ (Equation \eqref{eq:l_khz}), the confidence-interval Fisher randomization test (FRTCI) in \citet{ding2016randomization}, the martingale transformation test (MT) in \citet{CHUNG2021148}, and the subsampling quantile regression test (Subs) in \cite{chernozhukov2005subsampling}. Replicated potential outcomes follow a non-constant treatment effect model \citep{koenker2002inference,chernozhukov2005subsampling,ding2016randomization}, and specified by the equations $Y_i(0)=\varepsilon_i$, $Y_i(1)=1+Y_i(0)+\sigma_\tau Y_i(0)$ with $\varepsilon_i$ distributed standard exponential, log-normal, standard normal and $t_5$; half of the units were randomly assigned to treatment. Large values of $\sigma_\tau$ represent models with higher treatment effect heterogeneity.}
    \end{tablenotes}
    \end{ThreePartTable}
\end{table}


\begin{table}
\caption{\label{tab:power_adjusted} Size-adjusted power of 5\%-level tests when $H_0^\text{nocov}$
      is false (in percentage points)}
\centering
    \begin{ThreePartTable}
\begin{tabular}[t]{rrrrrrrrrrr}
\toprule
\multicolumn{1}{c}{$n$} & \multicolumn{5}{c}{$\sigma_\tau=0.2$} & \multicolumn{5}{c}{$\sigma_\tau=0.5$} \\
\cmidrule(l{3pt}r{3pt}){1-1} \cmidrule(l{3pt}r{3pt}){2-6} \cmidrule(l{3pt}r{3pt}){7-11}
\multicolumn{1}{c}{ } & \multicolumn{1}{c}{$L_{n,2}$} & \multicolumn{1}{c}{$L^{HKZ}$} & \multicolumn{1}{c}{Subs} & \multicolumn{1}{c}{FRTCI} & \multicolumn{1}{c}{MT} & \multicolumn{1}{c}{$L_{n,2}$} & \multicolumn{1}{c}{$L^{HKZ}$} & \multicolumn{1}{c}{Subs} & \multicolumn{1}{c}{FRTCI} & \multicolumn{1}{c}{MT} \\
\midrule
\addlinespace[0.3em]
\multicolumn{11}{l}{\textbf{Panel A: Exponential outcomes}}\\
\hspace{1em}50 & 8.3 & 9.5 & 7.2 & 6.2 & 7.6 & 20.4 & 25.5 & 19.5 & 16.8 & 11.0\\
\hspace{1em}100 & 12.5 & 13.4 & 12.4 & 12.1 & 9.0 & 45.8 & 49.1 & 34.6 & 41.7 & 14.8\\
\hspace{1em}400 & 42.0 & 36.3 & 24.4 & 35.8 & 12.1 & 94.4 & 94.7 & 90.4 & 95.3 & 23.4\\
\hspace{1em}800 & 61.4 & 63.0 & 47.8 & 70.0 & 16.6 & 100.0 & 100.0 & 99.7 & 99.9 & 32.3\\
\hspace{1em}1000 & 75.7 & 75.9 & 62.4 & 79.9 & 15.5 & 100.0 & 100.0 & 99.9 & 100.0 & 32.1\\
\addlinespace[0.3em]
\multicolumn{11}{l}{\textbf{Panel B: Log-normal outcomes}}\\
\hspace{1em}50 & 5.9 & 6.5 & 6.5 & 4.2 & 8.0 & 13.9 & 10.3 & 10.4 & 4.9 & 16.3\\
\hspace{1em}100 & 8.0 & 7.7 & 8.6 & 6.6 & 14.9 & 22.6 & 21.1 & 21.0 & 14.0 & 34.3\\
\hspace{1em}400 & 18.9 & 22.3 & 13.7 & 19.7 & 35.3 & 76.6 & 74.3 & 61.8 & 69.1 & 86.7\\
\hspace{1em}800 & 49.8 & 30.3 & 25.6 & 30.5 & 56.7 & 99.1 & 93.8 & 91.3 & 93.0 & 99.6\\
\hspace{1em}1000 & 56.5 & 41.5 & 33.4 & 40.3 & 64.5 & 99.8 & 98.2 & 96.3 & 97.6 & 99.7\\
\addlinespace[0.3em]
\multicolumn{11}{l}{\textbf{Panel C: Normal  outcomes}}\\
\hspace{1em}50 & 12.7 & 11.5 & 9.2 & 5.7 & 11.4 & 37.6 & 38.3 & 28.9 & 16.7 & 22.7\\
\hspace{1em}100 & 20.9 & 14.8 & 14.5 & 9.0 & 16.2 & 68.9 & 57.9 & 55.9 & 38.0 & 38.1\\
\hspace{1em}400 & 64.2 & 50.2 & 41.9 & 25.8 & 34.8 & 99.9 & 99.3 & 98.3 & 95.2 & 85.3\\
\hspace{1em}800 & 90.7 & 79.4 & 73.7 & 52.6 & 65.0 & 100.0 & 100.0 & 100.0 & 100.0 & 99.6\\
\hspace{1em}1000 & 95.7 & 88.6 & 84.6 & 69.7 & 78.8 & 100.0 & 100.0 & 100.0 & 100.0 & 100.0\\
\addlinespace[0.3em]
\multicolumn{11}{l}{\textbf{Panel D: $t_5$  outcomes}}\\
\hspace{1em}50 & 9.1 & 7.5 & 8.6 & 3.2 & 10.3 & 27.1 & 21.8 & 23.7 & 8.4 & 22.1\\
\hspace{1em}100 & 12.9 & 11.6 & 10.0 & 5.5 & 13.8 & 47.2 & 41.9 & 36.1 & 21.6 & 33.1\\
\hspace{1em}400 & 52.0 & 34.2 & 30.7 & 21.3 & 32.2 & 99.1 & 95.4 & 93.7 & 85.0 & 81.3\\
\hspace{1em}800 & 79.3 & 60.0 & 51.2 & 37.1 & 59.3 & 100.0 & 100.0 & 99.6 & 99.3 & 98.4\\
\hspace{1em}1000 & 89.2 & 73.5 & 65.8 & 51.2 & 68.9 & 100.0 & 100.0 & 100.0 & 99.9 & 99.6\\
\bottomrule
\end{tabular}
    \begin{tablenotes}
        \item{\small Estimates are based on 5,000 replications, which imply a simulation standard error of approximately 0.3 percentage points. The table shows the simulated power for the confidence interval permutation implementation of the characteristic-function based statistic in this paper's $L_{n,2}$ (Equation \eqref{eq:l_mine}), and \citet{henze2005checking}'s $L_n^{HKZ}$ (Equation \eqref{eq:l_khz}), the confidence-interval Fisher randomization test (FRTCI) in \citet{ding2016randomization}, the martingale transformation test (MT) in \citet{CHUNG2021148}, and the subsampling quantile regression test (Subs) in \cite{chernozhukov2005subsampling}. To compute size-adjusted power, I compute 5\% critical values using the corresponding percentiles of the simulated test statistic for each distribution and sample size across 5,000 simulations when $\sigma_\tau=0$.
        Replicated potential outcomes follow a non-constant treatment effect model \citep{koenker2002inference,chernozhukov2005subsampling,ding2016randomization}, and specified by the equations $Y_i(0)=\varepsilon_i$, $Y_i(1)=1+Y_i(0)+\sigma_\tau Y_i(0)$ with $\varepsilon_i$ distributed standard exponential, log-normal, standard normal and $t_5$; half of the units were randomly assigned to treatment. Large values of $\sigma_\tau$ represent models with higher treatment effect heterogeneity.}
    \end{tablenotes}
    \end{ThreePartTable}
\end{table}


\subsection{Simulations including covariates}

To assess the validity and power of the test based on $\mathcal{D}_{n,w}$, I simulate a completely randomized experiment in which potential outcomes and treatment effects follow a linear model in covariates as in \citet{ding2019decomposing}.%
\footnote{Although the simulations in that paper are to show the properties of a test for systematic treatment effect variation, those simulations encompass a framework where both systematic and idiosyncratic treatment effect variation may coexist.}
Specifically, potential outcomes observe the following equations:
\begin{gather*}
    Y_i(0) = 0.3+0.2X_{1i}+0.3X_{2i}-0.4X_{3i}+0.8X_{4i}+u_i, \qquad u_i\sim \mathcal{N}(0,0.26^2)
    \\
    Y_i(1) = \tau_i+Y_i(0),
    \\
    \tau_i=\delta_i+\varepsilon_i,
\end{gather*}
with covariates distributed as follows $X_{1i}\sim \mathcal{N}(0,1)$, $X_{2i}\sim \mathrm{Bernoulli}(0.5)$, $X_{3i}\sim \mathrm{Bernoulli}(0.25)$ and $X_{4i}\sim \mathcal{N}(0,1)$; $i=1,\dots,n$. The terms $\delta_i$ and $\varepsilon_i$ configure four types of variation in treatment effects: $\tau_i$  can have either no systematic ($\delta_i=0.3$) or systematic variation ($\delta_i=0.2 + 0.1X_{1i} + 0.4X_{3i}$); and $\tau_i$ can have either no idiosyncratic ($\varepsilon_i=0$) or idiosyncratic variation ($\varepsilon_i=\mathcal{N}(0,0.2^2)$). In each replication, sixty percent of the units are randomly assigned to treatment.

When implementing the test based on $\mathcal{D}_{n,\theta}$ (Equation \eqref{eq:d_n_theta}), I use the density function of the spherical stable distribution with parameter $\theta=2$ as weight function $w(t)$, and rewrite $\mathcal{D}_{n,\theta}$ in the following way:
\begin{align}
    \mathcal{D}_{n,2} = \frac{1}{n_0^2}& \sum_{i=1}^{n}\sum_{j=1}^{n}(1-D_i)(1-D_j) \exp(-|\widehat{\varepsilon}_i-\widehat{\varepsilon}_j|^2)+\frac{1}{n_1^2} \sum_{i=1}^{n} \sum_{j=1}^{n}  D_iD_j\exp(-|\widehat{\varepsilon}_i-\widehat{\varepsilon}_j|^2)
    \notag
    \\
    &-\frac{2}{n_0n_1} \sum_{i=1}^{n}\sum_{j=1}^{n} (1-D_i)D_j\exp(-|\widehat{\varepsilon}_i-\widehat{\varepsilon}_j|^2).
\end{align}
I select $\theta=2$ as suggested in the literature, which corresponds to a standard normal weighting function. As stated in Section \ref{sec:cov}, $\widehat{\varepsilon}_i$ is the residual from fitting a linear model of the observed outcome $Y_{i}$ on $D_i$, $\mathbf{X}_i=(X_{1i},X_{2i},X_{3i},X_{4i})$ and $D_i\mathbf{X}_i$.

To compute critical values, I use a permutation test based on \citet{oja1987permutation}. Specifically, I implement the following steps:
\begin{enumerate}
    \item \label{it:1} draw a vector of treatment assignments $D_1^*,\dots,D_n^*$ following the treatment assignment rule;
    \item \label{it:2} calculate the observed outcomes as $Y_i^* = Y_i + (D_i^*-D_i)\mathbf{e}'\widehat{\boldsymbol{\beta}}_{(D,D\mathbf{X}')}$, $i=1,\dots,n$, where $\mathbf{e}$ is a $(K+1)$-vector of ones and $\widehat{\boldsymbol{\beta}}_{(D,D\mathbf{X}')}$ is a $(K+1)$-vector of estimated coefficients associated with $(D,D\mathbf{X}')$ in the regression of $Y$ on $(D,\mathbf{X}',D\mathbf{X}')$;
    \item \label{it:3} calculate the statistic $\mathcal{D}_{n,2}^*$ using the sample $(D_1^*,Y_1^*),\dots,(D_n^*,Y_1^*)$;
    \item repeat steps \ref{it:1}-\ref{it:3} a large number of times (i.e. 2,000) and reject $H_0$ if $\mathcal{D}_{n,2}$ computed on \linebreak $(D_1,Y_1),\dots,(D_n,Y_n)$ exceeds the empirical $95\%$-quantile of the permutation distribution of $\mathcal{D}_{n,2}^*$.
\end{enumerate}

Table \ref{tb:cov_power} shows the rejection rates for 5\%-level tests based on $\mathcal{D}_{n,2}$. In the presence of no idiosyncratic treatment effect variation (columns 3 and 5), the test rejects about 5\% of the replications indicating the test's asymptotic validity. The test achieves power for large samples, which are similar to the sample sizes in the simulations in \citet{ding2019decomposing}, as shown in columns 2 and 4.

\begin{table}

\caption{Rejection rates of 5\%-level tests based on $\mathcal{D}_{n,2}$ for models with covariates
      (percentage points)}
\centering
\begin{ThreePartTable}
\label{tb:cov_power}
\begin{tabular}[t]{rrrr>{\raggedleft\arraybackslash}p{1in}}
\toprule
\multicolumn{1}{c}{ } & \multicolumn{4}{c}{Type of Variation} \\
\cmidrule(l{3pt}r{3pt}){2-5}
$n$ & None & Idiosyncratic & Systematic & Systematic \&
 Idiosyncratic\\
\midrule
200 & 5.50 & 9.7 & 5.51 & 9.3\\
500 & 5.50 & 16.6 & 5.50 & 16.6\\
1000 & 4.75 & 27.8 & 4.69 & 28.2\\
2000 & 5.90 & 49.9 & 5.90 & 49.9\\
5000 & 4.45 & 89.4 & 4.45 & 89.4\\
\bottomrule
\end{tabular}
\begin{tablenotes}
\item{\small Estimates are based on 10,000 replications, which imply a simulation standard error of approximately 0.2 percentage points. Replicated potential outcomes follow a linear model in covariates, as in \citet{ding2019decomposing}, and specified by the equations $Y_i(0)=f(X_{1i},X_{2i},X_{3i},X_{4i})+u_i$, $Y_i(1)=\tau_i+Y_i(0)$, $\tau_i=\delta_i+\varepsilon_i$, with $f(X_{1i},X_{2i},X_{3i},X_{4i})= 0.3+0.2X_{1i}+0.3X_{2i}-0.4X_{3i}+0.8X_{4i}$, $u_i\sim \mathcal{N}(0,0.26^2)$, $X_{1i}\sim \mathcal{N}(0,1)$, $X_{2i}\sim \mathrm{Bernoulli}(0.5)$, $X_{3i}\sim \mathrm{Bernoulli}(0.25)$ and $X_{4i}\sim \mathcal{N}(0,1)$, $i=1,\dots,n$. 60\% of the units were randomly assigned to treatment. The four types of variation correspond to four different combinations of the terms $\delta_i$ and $\varepsilon_i$: the replications have either no systematic ($\delta_i=0.3$) or systematic variation ($\delta_i=0.2 + 0.1X_{1i} + 0.4X_{3i}$), and either no idiosyncratic ($\varepsilon_i=0$) or idiosyncratic variation ($\varepsilon_i=\mathcal{N}(0,0.2^2)$).
}
\end{tablenotes}
\end{ThreePartTable}
\end{table}


\section{Application}

I apply the tests to the microcredit RCT implemented by \citet{augsburg2015impacts} in Bosnia and Herzegovina. The experiment targeted 1,196 loan applicants who were marginally rejected by an microfinance institution. The researchers offered the loan to 628 randomly selected applicants. Successful applicants received the loan within a week with similar interest rate and maturity to the regular institution's clients. The authors interviewed participants 14 months after. For those participants, the researchers collected baseline characteristics such as gender, age, marital status of the marginal applicant, and information on the household composition (number of children in the age ranges of 0--5, 6--10, 11--16, and number of household members that are: female, employed, in school, and retired). They found evidence that the treatment induced higher self-employment, increases in inventory, a reduction in the incidence of wage work, an increase in the labor supply of 16--19-year-olds in the household’s business, increases in profits and a reduction in consumption and savings. 

Table \ref{tab:app} shows the results for the characteristic-function based tests in 40 outcomes presented in the main tables in \citet{augsburg2015impacts}.\footnote{It is also possible to evaluate the multiple hypothesis that all 40 outcomes (or subsets of these outcomes) have no idiosyncratic treatment effect variation as pointed out above at the end of section \ref{sec:asym_beha}.} The table shows the empirical cumulative probability, or cumulative relative frequency, which is the fraction of quantiles of the corresponding resampling distribution that are below the calculated value of the statistic, of three statistics: $L_{n,2}$, $L_{n}^{HKZ}$ and $\mathcal{D}_{n}$. The resampling distributions are computed using 2000 samples with replacement (bootstrap) and permuting the treatment without replacement (permutation; see Section \ref{sec:simul} for details). The test based on $L_{n,2}$ rejects at the 5\% level if the empirical cumulative probability is lower than 2.5\% or greater than  97.5\%. The tests based on $L_{n}^{HKZ}$ and $\mathcal{D}_{n,2}$ reject at the 5\% level if the empirical cumulative probability is greater than 95\%.

In the case without covariates, the bootstrap and permutation tests based on the statistic $L_{n,2}$ reject the null of no treatment effect variation at 5\% significance level for 13 and 22 outcomes, respectively, whereas the test based on $L_{n}^{HKZ}$ rejects the null for 31 outcomes.

As expected, accounting for covariates tends to reject less often with only 12 outcomes rejecting the null based on $\mathcal{D}_{n,2}$. Likewise, as one would anticipate, those outcomes affected by the treatment are the candidates to exhibit treatment effect variation. Indeed, some of the outcomes for which the tests based on $L_{n,2}$ and $\mathcal{D}_{n,2}$ reject the null hypothesis of no idiosyncratic treatment effect coincide with the outcomes for which the authors find significant average treatment effects; those outcomes are related to loan take-up, ownership of inventory and hours worked by 16--19-year-olds in the household business. Additionally, both tests also reject the null of no idiosyncratic treatment effect for the following outcomes: self-employment income, an indicator for having 16--19-year olds, the number of 16--19-year olds in the household, and the home durable good index. In contrast to what \citet{augsburg2015impacts} state, the tests suggest that the distributions of the outcomes related to the presence of 16--19-year olds in the household do not seem to be the same across treatment and control groups.

The tests suggest that, for four outcomes, any treatment effect variation seems to be accounted for by the covariates collected by the researchers. Specifically, four (ten) outcomes are rejected by the bootstrap (permutation) test based on $L_{n,2}$ but not by the test based on $\mathcal{D}_{n,2}$, namely, the average yearly expenses of the main business, indicators for whether the household received self-employment income and wages, and the number of hours worked per household member in a week in other activities. 

Furthermore, Table \ref{tab:app_ext} shows the $p$-values for the confidence-interval Fisher randomization test (FRTCI) and the plug-in Fisher randomization test (FRTPI) in \citet{ding2016randomization}, the martingale transformation test in \citet{CHUNG2021148}, and the subsampling quantile regression test in \cite{chernozhukov2005subsampling} for each outcome in \citet{augsburg2015impacts}. The FRTCI is conservative in general detecting heterogeneity for only 3 outcomes, whereas the FRTPI and the subsampling tests detect heterogeneity in 18 and 16 outcomes, respectively. In principle, the FRT and the martingale permutation tests cannot be applied to a variety of outcomes in this application since they tend to be discrete.\footnote{\citet[][p. 669]{ding2016randomization} recognize that the Fisher randomization test does not directly address the case of discrete outcomes.}

\begin{landscape}
    
    \linespread{1}\small
    \centering
    
    \begin{ThreePartTable}
    \begin{TableNotes}[flushleft]
    \footnotesize \linespread{1}\small
    \item \footnotesize The table shows empirical cumulative probabilities for each of the statistics $L_{n,2}$, $L_{n}^{HKZ}$ and $\mathcal{D}_{n,2}$ and outcomes in \citet{augsburg2015impacts}.
    The empirical cumulative probability, or cumulative relative frequency, is the fraction of quantiles of the corresponding resampling (either bootstrap or permutation) distribution that are below the calculated value of the statistic. Resampling distributions are calculated using 2000 samples, which imply a simulation standard error of approximately 0.5 percentage points. The test based on $L_{n,2}$ rejects if the empirical cumulative probability is lower than 2.5\% or greater than  97.5\%. The tests based on $L_{n}^{HKZ}$ and $\mathcal{D}_{n,2}$ reject if the empirical cumulative probability is greater than 95\%. Covariates include age, separate indicators for gender, whether the respondent was married, divorced, and widowed; number of 5--11-year olds, number of 10--17-year olds, number of female HH members, number of HH members attending school, number of retired HH members, number of employed HH members. HH stands for household.
    \end{TableNotes}

        \begin{longtable}[t]{>{\raggedright\arraybackslash}p{4.0in}>{\raggedleft\arraybackslash}p{1.0in}>{\raggedleft\arraybackslash}p{1.0in}>{\raggedleft\arraybackslash}p{1.0in}>{\raggedleft\arraybackslash}p{1.2in}}
        \caption{\label{tab:app}Statistics' Empirical Cumulative Probability, in Percentage Points, for Tests for Idiosyncratic Treatment Effect on Outcomes
                      from the Microcredit Experiment in \citet{augsburg2015impacts}}\\
        \toprule
        \multicolumn{1}{c}{ } & \multicolumn{3}{c}{Without covariates} & \multicolumn{1}{c}{With covariates} \\
        \cmidrule(l{3pt}r{3pt}){2-4} \cmidrule(l{3pt}r{3pt}){5-5}
        Outcome & $L_{n,2}$ (Bootstrap) & $L_{n,2}$ (Permutation) & $L_{n}^{HKZ}$ (Permutation) & $\mathcal{D}_{n,2}$ (Permutation)\\
        \midrule
        \endfirsthead
        \caption[]{Statistics' Empirical Cumulative Probability, in Percentage Points \textit{(continued)}}\\
        \toprule
        \multicolumn{1}{c}{ } & \multicolumn{3}{c}{Without covariates} & \multicolumn{1}{c}{With covariates} \\
        \cmidrule(l{3pt}r{3pt}){2-4} \cmidrule(l{3pt}r{3pt}){5-5}
        Outcome & $L_{n,2}$ (Bootstrap) & $L_{n,2}$ (Permutation) & $L_{n}^{HKZ}$ (Permutation) & $\mathcal{D}_{n,2}$ (Permutation)\\
        \midrule
        \endhead
        \hline
        \insertTableNotes
        \endfoot
        \addlinespace[0.3em]
        \multicolumn{5}{l}{\textbf{Credit Outstanding at Endline}}\\
        \hspace{1em}Any loan outstanding (yes = 1) & 0.00 & 0.00 & 100.0 & 100.0\\
        \hspace{1em}Number of loans & 49.30 & 50.10 & 98.9 & 79.5\\
        \hspace{1em}One of outstanding loans is with bank (yes = 1) & 0.00 & 0.00 & 100.0 & 100.0\\
        \hspace{1em}At least one loan outstanding from a bank (yes = 1) & 0.05 & 0.00 & 100.0 & 100.0\\
        \addlinespace[0.3em]
        \multicolumn{5}{l}{\textbf{Self-Employment Activities: Revenues, Assets, and Profits}}\\
        \hspace{1em}Value of all assets owned & 55.40 & 69.45 & 100.0 & 17.2\\
        \hspace{1em}Ownership of inventory (yes = 1) & 99.15 & 99.20 & 98.7 & 99.1\\
        \hspace{1em}Average yearly revenue of main business & 95.20 & 99.95 & 100.0 & 45.0\\
        \hspace{1em}Average yearly expenses of main business & 97.40 & 100.00 & 100.0 & 2.9\\
        \hspace{1em}Average yearly profit of main business & 96.50 & 99.95 & 100.0 & 98.7\\
        \hspace{1em}Respondent owns a business (yes = 1) & 5.65 & 3.45 & 96.0 & 94.5\\
        \hspace{1em}Main business is in service (yes = 1) & 92.00 & 91.35 & 83.5 & 79.9\\
        \hspace{1em}Main business is in agriculture/farming (yes = 1) & 91.65 & 92.30 & 83.8 & 84.2\\
        \hspace{1em}HH has started a business (since baseline) (yes = 1) & 84.20 & 82.25 & 67.4 & 67.9\\
        \hspace{1em}HH has closed their business (since baseline) (yes = 1) & 16.20 & 17.05 & 65.7 & 63.1\\
        \addlinespace[0.3em]
        \multicolumn{5}{l}{\textbf{Income}}\\
        \hspace{1em}Income from self-employment received (yes = 1) & 1.45 & 1.65 & 96.8 & 94.0\\
        \hspace{1em}Amount of income from self-employment received (yearly) & 100.00 & 100.00 & 100.0 & 98.2\\
        \hspace{1em}Income from wages received (yes = 1) & 99.15 & 99.10 & 99.0 & 94.5\\
        \hspace{1em}Income from wages received (yearly) & 45.80 & 41.20 & 100.0 & 97.1\\
        \hspace{1em}Income from remittances received (yes = 1) & 17.65 & 17.25 & 64.7 & 56.2\\
        \hspace{1em}Amount of income from remittances received (yearly) & 50.65 & 52.10 & 100.0 & 14.6\\
        \hspace{1em}Income from government benefits received (yes = 1) & 5.40 & 6.35 & 88.7 & 66.5\\
        \hspace{1em}Amount of income from government benefits received (yearly) & 36.55 & 23.15 & 100.0 & 88.5\\
        \addlinespace[0.3em]
        \multicolumn{5}{l}{\textbf{Consumption and Savings}}\\
        \hspace{1em}Total consumption per capita & 27.50 & 7.70 & 98.3 & 56.8\\
        \hspace{1em}Amount spent on durable consumption in last year & 7.10 & 0.25 & 100.0 & 65.5\\
        \hspace{1em}Nondurable consumption & 12.05 & 1.65 & 100.0 & 52.0\\
        \hspace{1em}Total food consumption in last week & 88.80 & 99.50 & 100.0 & 60.5\\
        \hspace{1em}Amount spent on education in last year & 72.20 & 87.75 & 100.0 & 37.9\\
        \hspace{1em}Amount spent on alcohol, cigarettes, tobacco in last week & 3.90 & 0.00 & 100.0 & 75.9\\
        \hspace{1em}Recreation & 23.70 & 9.10 & 100.0 & 60.8\\
        \hspace{1em}Home durable good index & 0.10 & 0.10 & 100.0 & 100.0\\
        \hspace{1em}Estimated amount of savings & 8.95 & 0.50 & 100.0 & 100.0\\
        \addlinespace[0.3em]
        \multicolumn{5}{l}{\textbf{Time Worked by Household Members}}\\
        \hspace{1em}Hours worked per HH member in a week, total & 5.65 & 3.55 & 10.8 & 21.1\\
        \hspace{1em}Hours worked per HH member in a week, business & 92.70 & 99.85 & 100.0 & 95.8\\
        \hspace{1em}Hours worked per HH member in a week, other activities & 0.20 & 0.00 & 100.0 & 42.4\\
        \hspace{1em}Hours worked per HH member in a week, total, teens & 97.65 & 98.35 & 98.0 & 79.3\\
        \hspace{1em}Hours worked per HH member in a week, business, teens & 98.15 & 99.25 & 100.0 & 99.8\\
        \hspace{1em}Hours worked per HH member in a week, other activities, teens & 61.60 & 66.80 & 11.6 & 69.4\\
        \addlinespace[0.3em]
        \multicolumn{5}{l}{\textbf{Social Impacts}}\\
        \hspace{1em}Stress Score & 21.05 & 19.45 & 59.2 & 81.5\\
        \hspace{1em}Having kids in the age range 16-19 & 100.00 & 100.00 & 99.9 & 96.2\\
        \hspace{1em}Number of kids 16-19 & 100.00 & 100.00 & 100.0 & 99.3\\*
        \end{longtable}

    \end{ThreePartTable}

\end{landscape}

\section{Conclusion}

I show the usefulness of the tests based on the characteristic function to detect idiosyncratic treatment effect variation. The tests are asymptotically valid and have power against plausible alternatives. As further variation is detected, the test can help researchers in their data collection efforts to determine who benefit the most out of policy interventions, to disentangle causal mechanisms, and to identify characteristics that predict treatment effect variation. An application of the tests to a microcredit experiment in Bosnia and Herzegovina shows that outcomes related to loan take-up, self-employment, savings, expenses on durable goods, and hours worked by 16--19-year-olds in the household business exhibit idiosyncratic treatment effect variation not accounted for baseline characteristics. For those outcomes, researchers could potentially try to collect more covariates to inspect the remaining treatment effect heterogeneity, and potentially, improve treatment targeting.

\newpage
\bibliography{biblio} 

\appendix

\newpage
\setcounter{table}{0}
\renewcommand{\thetable}{A\arabic{table}}

\newpage


\section{Proofs}
\label{app:proofs}

\subsection{Proof of Proposition \ref{prop:l_w}}
\begin{proof}[\unskip\nopunct]
The proposition directly follows from the arguments laid out in Equations \eqref{h0:noCovCF}-\eqref{eq:l_w}; so under $H_{0}^\text{nocov}$, $L_w=0$.
\end{proof}

\subsection{Proof of Corollary \ref{co:l_theta}}
\begin{proof}[\unskip\nopunct]
Let's start by considering $L_w$ from Equation \eqref{eq:l_w},
\begin{align*}
    L_w & = \int_{\mathbb{R}} \left\{ |\varphi_1(t)|^2-|\varphi_0(t)|^2 \right\} w(t) dt,
    \\
    & = \int_{\mathbb{R}} \EX\left[\cos\{t[Y(1)-Y'(1)]\}-\cos\{t[Y(0)-Y'(0)]\}\right]  w(t) dt,
    \\
    & = \EX\left[\int_{\mathbb{R}} (\cos\{t[Y(1)-Y'(1)]\}-\cos\{t[Y(0)-Y'(0)]\})  w(t) dt\right].
\end{align*}    
    The first equality follows from the definition of $L_w$ in Equation \eqref{eq:l_w}. The second equality follows from applying the fact that $|\varphi_W(t)|^2 = \EX[\cos \{t(W-W')\}]$, with $W'$ being an independent variable with the same distribution as $W$, to both $Y(1)$ and $Y(0)$  \citep[see][p.\ 179]{szekely2005hierarchical}. The third equality follows by Fubini's theorem.
    
    Now, let $w_\theta(t)$ be the density of a spherical stable distribution with parameter $\theta$, and let's define $L_\theta$ as
\begin{align*}    
    L_\theta & = \EX\left[\int_{\mathbb{R}} (\cos\{t[Y(1)-Y'(1)]\}-\cos\{t[Y(0)-Y'(0)]\})  w_\theta(t) dt\right],
    \\
    & = \EX\left[ \exp\{-|Y(1)-Y'(1)|^\theta\}-\exp\{-|Y(0)-Y'(0)|^\theta\} \right].
\end{align*}
 The equality follows from the following property of the spherical stable distribution with parameter $\theta$:
\begin{gather*}
    \int_\mathbb{R} \cos\{tX\} w_\theta(t)dt = \exp(-|X|^\theta).
\end{gather*}
\end{proof}


\subsection{Proof of Theorem \ref{thm:l_n}}

\begin{proof}[\unskip\nopunct]
The proof is based on \citet{meintanis2005permutation} and \citet{chen2019some}. Note that
\begin{align}
     \varphi_{n,1}(t)  & = \frac{1}{n_1} \sum_{j=1}^{n} D_i 
    \left[\cos\{tY_j\}+\mathrm{i}\sin\{tY_j\}\right],
    \notag
    \\
    & = \frac{1}{n_1} \sum_{j=1}^{n_1} 
    \left[\cos\{tY_{j}(1)\}+\mathrm{i}\sin\{tY_{j}(1)\}\right],
    \notag
    \\
   \varphi_{n,1}(t) - \varphi_{1}(t)  & = \frac{1}{n_1} \sum_{j=1}^{n_1} 
    \left[\cos\{tY_{j}(1)\}+\mathrm{i}\sin\{tY_{j}(1)\}-\varphi_{1}(t)\right],
    \notag
    \\
    \varphi_{n,1}(t) - \varphi_{1}(t)  & = \frac{1}{n_1}\sum_{j=1}^{n_1} h(Y_j(1),t),
    \notag
\end{align}
where $h(Y_1(1),t),\dots,h(Y_{n_1}(1),t)$ and $h(Y_1(1),t),\dots,h(Y_{n_1}(1),t)$ are zero-mean independent and identically distributed random elements of $\mathcal{L}_2$, $\EX[\norm{h(Y_1(1),t)}_w^2]<\infty$ and \linebreak $\EX[\norm{h(Y_1(0),t)}_w^2]<\infty$ (notice that $|h(W,t)|<1$ for $W=Y_1(0),Y_1(1)$ and for all $t\in\mathbb{R}$). Note that this process involves continuous cosine and sine functions and thus is continuous in the Hilbert space. 

Let $ Z_{n,1}(t)\equiv \varphi_{n,1}(t) - \varphi_{1}(t)$ and $ Z_{n,0}(t)\equiv \varphi_{n,0}(t) - \varphi_{0}(t)$. Then, by the Hilbert space Central Limit Theorem \citep[see][Section 1.8]{van1996weak}, as $n \rightarrow \infty$,
\begin{gather}
    \sqrt{\frac{n_0n_1}{n}}Z_{n,1}(t) \rightsquigarrow \pi_0^{1/2} Z_1(t).
    \label{eq:z1}
\end{gather}
where $\{Z_1(t):t\in \mathbb{R}\}$ is a zero-mean complex Gaussian process with covariance function given, for all $t,s \in \mathbb{R}$, by
\begin{align*}
    \gamma_1(s,t) & = \EX[Z_1(s)\overline{Z_1(t)}]
     = \varphi_1(s-t)-\varphi_1(s)\varphi_1(-t),
\end{align*}
and complementary covariance function given, for all $s,t\in \mathbb{R}$, by
\begin{align*}
    \gamma_1^C(s,t) & = \EX[Z_1(s)Z_1(t)\}] =\varphi_1(s+t)-\varphi_1(s)\varphi_1(t).
\end{align*}
Likewise,
\begin{gather}
    \sqrt{\frac{n_0n_1}{n}}Z_{n,0}(t)\rightsquigarrow \pi_1^{1/2} Z_0(t),
    \label{eq:z0}
\end{gather}
where $\{Z_0(t):t\in \mathbb{R}\}$ is a zero-mean complex Gaussian process with covariance function given, for all $t,s \in \mathbb{R}$, by
\begin{align*}
    \gamma_0(s,t) & = \EX[Z_0(s)\overline{Z_0(t)}]
     = \varphi_0(s-t)-\varphi_0(s)\varphi_0(-t)    ,
\end{align*}
and complementary covariance function given, for all $s,t\in \mathbb{R}$, by
\begin{align*}
    \gamma_0^C(s,t) & = \EX[Z_0(s)Z_0(t)]
    = \varphi_0(s+t)-\varphi_0(s)\varphi_0(t).
\end{align*}
Now,
\begin{align*}
    \frac{n_0n_1}{n}\left[|\varphi_{n,1}(t)|^2-|\varphi_{n,0}(t)|^2 \right]
    & = \frac{n_0n_1}{n}\left[|Z_{n,1}(t)+\varphi_{1}(t)|^2-|Z_{n,0}(t)+\varphi_{0}(t)|^2\right], 
    \\
    & = \frac{n_0n_1}{n} \left\{|Z_{n,1}(t)|^2-|Z_{n,0}(t)|^2+|\varphi_{1}(t)|^2-|\varphi_{0}(t)|^2\right.
    \\
    & \phantom{ho} +\Real[\varphi_1(t)]\Real[Z_{n,1}(t)]+\Imag[\varphi_1(t)]\Imag[Z_{n,1}(t)]
    \\
    & \phantom{ho}\left. -\Real[\varphi_0(t)]\Real[Z_{n,0}(t)]+\Imag[\varphi_0(t)]\Imag[Z_{n,0}(t)]\right\}.
    \\
    & = \frac{n_0n_1}{n} \left\{|Z_{n,1}(t)|^2-|Z_{n,0}(t)|^2\right.
    \\
    & \phantom{ho} +\Real[\varphi_1(t)]\Real[\varphi_{n,1}(t)]+\Imag[\varphi_1(t)]\Imag[\varphi_{n,1}(t)]
    \\
    & \phantom{ho}\left. -\Real[\varphi_0(t)]\Real[\varphi_{n,0}(t)]+\Imag[\varphi_0(t)]\Imag[\varphi_{n,0}(t)]\right\}.    
\end{align*}
The first equality follows from adding and subtracting the terms $\varphi_{1}(t)$ and $\varphi_{0}(t)$, and by the definitions of $Z_{n,1}(t)$ and $Z_{n,0}(t)$. The second equality follows from straightforward algebra. The third equality follows from the definitions of $Z_{n,1}(t)$ and $Z_{n,0}(t)$, and the fact that $|a|^2=\Real^2(a)+\Imag^2(a)$ for a complex number $a$.
Thus,
\begin{align}
    \frac{n_0n_1}{n}\left\{|\varphi_{n,1}(t)|^2-|\varphi_{n,0}(t)|^2-\Delta^2_{n}(t) \right\}
    & = \frac{n_0n_1}{n}\left[|Z_{n,1}(t)|^2-|Z_{n,0}(t)|^2\right], 
    \label{eq:delta2t}
\end{align}
where
\begin{gather*}
    \Delta^2_{n}(t) = \Delta^2_{n,1}(t) - \Delta^2_{n,0}(t),
    \\
    \Delta^2_{n,1}(t) = \Real[\varphi_1(t)]\Real[\varphi_{n,1}(t)]+\Imag[\varphi_1(t)]\Imag[\varphi_{n,1}(t)],
    \\
    \Delta^2_{n,0}(t)= \Real[\varphi_0(t)]\Real[\varphi_{n,0}(t)]+\Imag[\varphi_0(t)]\Imag[\varphi_{n,0}(t)].   
\end{gather*}
Then, as $n\rightarrow \infty$, for all $t\in \mathbb{R}$, by the continuous mapping theorem \citep[see][Theorem 1.3.6]{van1996weak} and Equations \eqref{eq:z1} and \eqref{eq:z0},
\begin{gather*}
    \frac{n_0n_1}{n}\left[|Z_{n,1}(t)|^2-|Z_{n,0}(t)|^2\right]  \rightsquigarrow \pi_0|Z_1(t)|^2 -  \pi_1|Z_0(t)|^2.
\end{gather*}
Since 
\begin{align*}
    \frac{n_0n_1}{n}\left[L_{n,w} -\int_\mathbb{R}\Delta^2_{n}(t)w(t)dt\right]
    &= \frac{n_0n_1}{n}\int_\mathbb{R}\left[|\varphi_{n,1}(t)|^2-|\varphi_{n,0}(t)|^2 -\Delta^2_{n}(t)\right]w(t)dt,
    \\
    &=\frac{n_0n_1}{n}\int_\mathbb{R}\left[|Z_{n,1}(t)|^2-|Z_{n,0}(t)|^2\right]w(t)dt,
\end{align*}
where the first equality follows from the definitions of $L_{n,w}$ and $\norm{\cdot}_w$, and Equation \eqref{eq:delta2t}. By invoking the continuous mapping theorem \citep[see][Theorem 1.3.6]{van1996weak}, we find that, as $n\rightarrow \infty$,
\begin{align*}
    \frac{n_0n_1}{n}\left[L_{n,w} -\int_\mathbb{R}\Delta^2_{n}(t)w(t)dt\right] &\rightsquigarrow 
    \int_\mathbb{R} [\pi_0|Z_1(t)|^2 -  \pi_1|Z_0(t)|^2] w(t) dt.
\end{align*}

\end{proof}


\subsection{Proof of Theorem \ref{thm:l_n_H0}}

\begin{proof}[\unskip\nopunct]
Note that
\begin{align}
    |\varphi_{n,1}(t)|^2-|\varphi_{1}(t)|^2 & = \frac{1}{n_1^2} \sum_{j,l=1}^{n_1} 
    \left[\cos\{t[Y_{j}(1)-Y_{l}(1)]\}\right] - \EX[\cos\{t[Y(1)-Y'(1)]\}],
    \notag
    \\
    & = \frac{1}{n_1^2} \sum_{j,l=1}^{n_1} 
    \left[\cos\{t[Y_{j}(1)-Y_{l}(1)]\} - \EX[\cos\{t[Y(1)-Y'(1)]\}]\right],
    \notag
    \\
    & = \frac{1}{n_1^2} \sum_{j\neq l}^{n_1} 
    h(Y_{j}(1)-Y_{l}(1),t) + \frac{1}{n_1}\{1-\EX[\cos\{t[Y(1)-Y'(1)]\}]\},
    \notag
    \\
    & = \tfrac{n_1-1}{n_1} \frac{1}{\binom{n}{2}} \sum_{j<l}^{n_1} 
    h(Y_{j}(1)-Y_{l}(1),t) + \frac{1}{n_1}\{1-\EX[\cos\{t[Y(1)-Y'(1)]\}]\}.
    \label{eq:diff_varphi1}
\end{align}
The first equality follows from straightforward algebra and the fact that $|\varphi_{1}(t)|^2 = \EX[\cos\{t[Y(1)-Y'(1)]\}]$, where $Y'(1)$ is a independent random variable with the same distribution as $Y(1)$ \citep[see][p.\ 179]{szekely2005hierarchical}. The third equality follows from the fact that $\cos(0)=1$ and $h(Y_{j}(1)-Y_{l}(1),t)=\cos\{t[Y_{j}(1)-Y_{l}(1)]\} - \EX[\cos\{t[Y(1)-Y'(1)]$. 

Note that $h(Y_{j}(1)-Y_{l}(1),t)$ is an element of the Hilbert space and a symmetric kernel of degree 2 with $\EX[h(\cdot,t)^2]<\infty$ ($|h(\cdot,t)^2|<2$ for all $t$). Therefore, by the Hilbert space Central Limit Theorem (see \citealp[Section 1.8]{van1996weak}; \citealp[Theorem 5.5.1.A]{serfling2009approximation}), as $n \rightarrow \infty$,    
\begin{gather}
    \sqrt{n_1}\frac{2}{n_1(n_1-1)} \sum_{j<l}^{n_1} 
    h(Y_{j}(1)-Y_{l}(1),t) \rightsquigarrow \zeta_1(t),
    \label{eq:llaw_ustat}
\end{gather}
with $\{\zeta_1(t):t\in \mathbb{R}\}$ a zero-mean Gaussian process with covariance function given, for all $s,t\in \mathbb{R}$, by
\begin{multline*}
    \EX[\zeta_1(t)\zeta_1(s)]=\mathrm{Cov}\{\Real(\varphi_1(t)) \cos(tY(1))+\Imag(\varphi_1(t)) \sin(tY(1)),\\\Real(\varphi_1(s)) \cos(sY(1))+\Imag(\varphi_1(s)) \sin(sY(1))\}.    
\end{multline*}
Then, from Equations \eqref{eq:diff_varphi1} and \eqref{eq:llaw_ustat},
\begin{gather}
    \sqrt{\frac{n_0n_1}{n}}\left[|\varphi_{n,1}(t)|^2-|\varphi_{1}(t)|^2 \right] \rightsquigarrow \pi_0^{1/2} \zeta_1(t), 
\end{gather}
and, similarly,
\begin{gather}
    \sqrt{\frac{n_0n_1}{n}}\left[|\varphi_{n,0}(t)|^2-|\varphi_{0}(t)|^2 \right] \rightsquigarrow \pi_1^{1/2} \zeta_0(t), 
\end{gather}
with $\{\zeta_0(t):t\in \mathbb{R}\}$ a zero-mean Gaussian process with the same distribution as $\{\zeta_1(t):t\in \mathbb{R}\}$ under $H_{0}^\text{nocov}$. 
Since 
\begin{gather}
    \sqrt{\frac{n_0n_1}{n}}\left[L_{n,w}-\norm{\varphi_1(t)}_w^2+\norm{\varphi_0(t)}_w^2\right] = \int_\mathbb{R}\sqrt{\frac{n_0n_1}{n}}\left[|\varphi_{n,1}(t)|^2-|\varphi_{1}(t)|^2 -|\varphi_{n,0}(t)|^2+|\varphi_{0}(t)|^2 \right]w(t)dt.
    \label{eq:proof_3}
\end{gather}
One can write the term on the left side of Equation \eqref{eq:proof_3} more succinctly under  $H_0^{\text{nocov}}$; since $\varphi_1(t)=\varphi_0(t)$ for all $t\in \mathbb{R}$,  $\norm{\varphi_1(t)}_w^2=\norm{\varphi_0(t)}_w^2$. Finally, by invoking the continuous mapping theorem, we find that, as $n\rightarrow \infty$,
\begin{gather*}
    \sqrt{\frac{n_0n_1}{n}}L_{n,w} \rightsquigarrow \int_\mathbb{R}\left[\pi_0^{1/2}\zeta_1(t)-\pi_1^{1/2} \zeta_0(t)\right]w(t)dt.
\end{gather*}
\end{proof}


\subsection{Proof of Proposition \ref{prop:d_w}}

\begin{proof}[\unskip\nopunct]
Assumption \ref{asm:cef} guarantees that the random variables $\varepsilon(1)=Y(1) - \EX[Y(1)|\mathbf{X}]$ and $\varepsilon(0)=Y(0) - \EX[Y(0)|\mathbf{X}]$ exist. The rest of the proof follows from Proposition 1 in \citet{chen2019some}.
\end{proof}


\subsection{Auxiliary lemmas required for the proof of Theorems \ref{thm:d_w_stoch_lim} and \ref{thm:d_w_asym}}

Let $\varphi_{\widetilde{\varepsilon}(d)}(t) = \frac{1}{n_d}\sum_{j:D_j=d}^{n}\exp\{\mathrm{i}t\varepsilon_j(d)\}$.

\begin{lemma}
    \label{le:res_taylor}
    For $d=0,1$,
    \begin{gather}
        \varphi_{\widehat{\varepsilon}(d)}(t)=\varphi_{\widetilde{\varepsilon}(d)}(t)+\mathrm{i}\frac{t}{n_d}\sum_{j:D_j=d}^{n}\exp\{\mathrm{i}t\varepsilon_j(d)\}\left[\widehat{m}_d(X_j)-m_d(X_j)\right]+t^2r_d(t)
    \end{gather}
    with $\sup_t|r_d(t)|=o_P(n^{-1/2})$.
\end{lemma}
\begin{proof}
    The proof is adapted from Lemma 10(i) in \citet{pardo2015tests}. For $d=0,1$,
    \begin{align*}
        \varphi_{\widehat{\varepsilon}(d)}(t)
        &=\frac{1}{n_d}\sum_{j:D_j=d}^{n}\exp\{\mathrm{i}t\widehat{\varepsilon}_j(d)\},
        \\
        &=\frac{1}{n_d}\sum_{j:D_j=d}^{n}\exp\{\mathrm{i}t\varepsilon_j(d)\} + \mathrm{i}\frac{t}{n_d}\sum_{j:D_j=d}^{n}\exp\{\mathrm{i}t\varepsilon_j(d)\}\left[\widehat{\varepsilon}_j(d)-\varepsilon_j(d)\right]
        \\
        & \phantom{\hspace{0.5cm}}+t^2R_d(t)\frac{1}{n_d}\sum_{j:D_j=d}^{n}\left[\widehat{\varepsilon}_j(d)-\varepsilon_j(d)\right]^2,
        \\
        &=\varphi_{\widetilde{\varepsilon}(d)}(t)+\frac{\mathrm{i}t}{n_d}\sum_{j:D_j=d}^{n}\exp\{\mathrm{i}t\varepsilon_j(d)\}\left[\widehat{m}_d(X_j)-m_d(X_j)\right]
        \\
        & \phantom{\hspace{0.5cm}}+t^2R_d(t)\frac{1}{n_d}\sum_{j:D_j=d}^{n}\left[\widehat{m}_d(X_j)-m_d(X_j)\right]^2,  
        \\
        &=\varphi_{\widetilde{\varepsilon}(d)}(t)+\frac{\mathrm{i}t}{n_d}\sum_{j:D_j=d}^{n}\exp\{\mathrm{i}t\varepsilon_j(d)\}\left[\widehat{m}_d(X_j)-m_d(X_j)\right]+t^2r_d(t).          
    \end{align*}
   The first equality follows from the definition of $\varphi_{\widehat{\varepsilon}(d)}(t)$. The second equality follows from using a Taylor expansion of $\exp\{\mathrm{i}t\widehat{\varepsilon}_j(d)\}$ around $\varepsilon_j(d)$ such that $\sup_t|R_d(t)|=O_P(1)$. The third equality follows from the definitions of $\varepsilon(d)$, $\widehat{\varepsilon}(d)$, and  $\varphi_{\widetilde{\varepsilon}(d)}(t)$. The last equality follows from $\sup_{x\in R}|\widehat{m}_d(x)-m_d(x)|=o_P(n_d^{-1/4})$ (see Equation \ref{eq:mx_lim}).
        
\end{proof}


Let $\varphi_{\pi_1}(t)\equiv\pi_1\varphi_1(t)+(1-\pi_1)\varphi_0(t)$ and $U_d^0(t)\equiv\sqrt{n_d}\left\{\varphi_{\widehat{\varepsilon}(d)}(t)-\varphi_{\pi_1}(t)\right\}$, $d=0,1$.

\begin{lemma}
    \label{le:d_w_fact}
    Under Assumptions \ref{asm:weight} and \ref{asm:kernel},
    \begin{gather}
        \frac{n_0n_1}{n}\mathcal{D}_{n,w}=\norm{Z_{n}}_w^2,
    \end{gather}
    with $Z_{n}(t)=\sqrt{\frac{n_0}{n}}U_1(t)-\sqrt{\frac{n_1}{n}}U_0(t)$, $U_d(t)=U_{0d}(t)+t^2\rho_d^2(t)$ and 
    \begin{multline*}
        U_{0d}(t)=\frac{1}{\sqrt{n_d}}\sum_{l=1}^{n_d}\left\{\cos[t\varepsilon_l(d)]+\sin[t\varepsilon_l(d)]+t\varepsilon_l(d)\{\Imag[\varphi_{\varepsilon(d)}(t)]-\Real[ \varphi_{\varepsilon(d)}(t)]\}\right.\\ \left.-\Real[\varphi_{\pi_1}(t)]-\Imag[\varphi_{\pi_1}(t)]\right\}
    \end{multline*}
    $\sup_t|\rho_d(t)|=o_P(n^{-1/2})$.
\end{lemma}


\begin{proof}
The proof is adapted from Lemma 1 in \citet{rivas2019two}. Note that
    \begin{align*}
         \frac{n_0n_1}{n}\mathcal{D}_{n,w}&=\norm{Z_n^0(t)}_w^2,
    \end{align*}
    where
    \begin{align*}
        Z_n^0(t) &= \sqrt{\frac{n_0n_1}{n}}\left[\varphi_{\widehat{\varepsilon}_1}(t)-\varphi_{\widehat{\varepsilon}_0}(t)\right],
        \\
        &= \sqrt{\frac{n_0n_1}{n}}\left[\varphi_{\widehat{\varepsilon}_1}(t)+\varphi_{\pi_1}(t)-\varphi_{\pi_1}(t)-\varphi_{\widehat{\varepsilon}_0}(t)\right],
        \\
        &=    \sqrt{\frac{n_0}{n}}U_1^0(t)-\sqrt{\frac{n_1}{n}}U_0^0(t).
    \end{align*}
    The first equality follows from the definition of $\mathcal{D}_{n,w}$. The second equality follows from adding and subtracting $\varphi_{\pi_1}(t)$. The third equality follows from the definition of $U_d^0(t)$, $d=0,1$.
    
    From Lemma \ref{le:res_taylor}, 
    \begin{gather*}
        U_d^0(t) = \sqrt{n_d}[\varphi_{\widetilde{\varepsilon}(d)}(t)-\varphi_{\pi_1}(t)]+A_{d,1}(t)+t^2\rho_{d,1}(t), \qquad d=0,1,
    \end{gather*}
    where $\sup_t|\rho_{d,1}(t)|=o_P(n^{-1/2})$ and
    \begin{gather*}
        A_{d,1}(t)=\frac{\mathrm{i}t}{\sqrt{n_d}}\sum_{j:D_j=d}^{n}\exp\{\mathrm{i}t\varepsilon_j(d)\}\left[\widehat{m}_d(X_j)-m_d(X_j)\right].
    \intertext{Moreover, from Equation (6) of the supplementary material for \citet{pardo2015nonparam},}
        A_{d,1}(t)=-\mathrm{i}t\varphi_d(t)\frac{1}{\sqrt{n_d}}\sum_{j:D_j=d}^{n}\varepsilon_j(d) + t\rho_{d,2}(t)
    \end{gather*}
    with $\sup_t|\rho_{d,2}(t)|=o_P(1)$.
    
    All of the above facts and Assumption \ref{asm:weight} imply that
    \begin{gather*}
        \norm{Z_{n}^0}_w^2 = \norm{Z_n}_w^2
    \end{gather*}
\end{proof}

\subsection{Proof of Theorem \ref{thm:d_w_stoch_lim}}

\begin{proof}[\unskip\nopunct]
The proof is adapted from the proof of Theorem 1 in \citet[][p. 1390]{rivas2019two}. Note that under Assumption \ref{asm:kernel} and following Theorem 8 of \citet{hansen2008uniform}
\begin{gather}
    \sup_{x\in S} |\widehat{m}_d(x)-m_d(x)|=O_P\left(\left(\frac{\ln n}{nh^q}\right)^{1/2}\right).
    \label{eq:mx_lim_pre1}
\end{gather}
Consequently, if $(\ln n/nh^q)^{1/2}\rightarrow 0$ as $n\rightarrow 0$, then
\begin{gather}
    \sup_{x\in S} |\widehat{m}_d(x)-m_d(x)|=o_P(1),
    \label{eq:mx_lim_pre2}
\end{gather}
which implies that
\begin{gather}
    \sup_{x\in S} |\widehat{m}_d(x)-m_d(x)|=o_P(n^{-1/4}).
    \label{eq:mx_lim}
\end{gather}

From Lemma \ref{le:res_taylor} and Equation \eqref{eq:mx_lim}
\begin{align*}
    \mathcal{D}_{n,w} & = \norm{\varphi_{\widetilde{\varepsilon}(1)}-\varphi_{\widetilde{\varepsilon}(0)}}_w^2+o_P(1).
    \\
    & = \norm{\varphi_{\varepsilon(1)}-\varphi_{\varepsilon(0)}}_w^2+o_P(1).
\end{align*}
The second equality follows from Theorem 2.3 in \citet{meintanis2005permutation}, namely, $\norm{\varphi_{\widetilde{\varepsilon}(1)}-\varphi_{\widetilde{\varepsilon}(0)}}_w^2\xrightarrow[]{P}\norm{\varphi_{\varepsilon(1)}-\varphi_{\varepsilon(0)}}_w^2$. 
\end{proof}

\subsection{Proof of Theorem \ref{thm:d_w_asym}}

\begin{proof}[\unskip\nopunct]
The proof is adapted from the proof of Theorem 2 in \citet{rivas2019two}. By the central limit theorem for independent and identically random elements in Hilbert spaces $\{U_{0d}(t),t\in\mathbb{R}\}$ converges to a zero-mean Gaussian process $U^{(d)}$ on $\mathcal{L}_2$ with covariance structure $\varrho_{0}(s,t)$, $d=0,1$. For constants $a$ and $b$ such that $a^2+b^2=1$, the centered process $Z^{\text{cov}}(t)=aU^{(1)}(t)+bU^{(0)}(t)$ has covariance structure $\varrho_{0}(s,t)$ and since $n_1/n\xrightarrow[]{P}\pi_1$, it follows that $\{Z^{\text{cov}}_n(t),t\in\mathbb{R}\}$ converges in law to $\{Z^{\text{cov}}(t),t\in\mathbb{R}\}$ under $H_0$. Finally, the result follows from the continuous mapping theorem.
    
\end{proof}


\subsection{Randomization inference properties of the permutation test of $L_{n,w}$ described in Section \ref{sec:simul}}

The proposition in this section states that the permutation distribution of statistic $L_{n,w}$ in Section \ref{sec:simul} is in fact a randomization inference test. As such, the test inherits the nice properties of randomization tests. 

Let $\mathbf{D} = (D_1,\dots,D_n)$ be a vector that stacks treatment assignments $D_i$, and $\mathcal{D}$ the set of all possible treatment assignments $\mathbf{D}$ under the treatment assignment rule. Let $\mathbf{D},\mathbf{D}^* \in \mathcal{D}$. For unit $i$, let 
\begin{gather*}
    Y_{i,\tau_0}^{RI}(D_i,D_i^*) = (1-D_i)Y_i(0)+D_iY_i(1)+\tau_0(D_i^*-D_i)    
\end{gather*}
be the permuted outcome under randomization inference when $i$'s original treatment assignment was $D_i$ and the permuted treatment assignment is $D_i^*$, and for a hypothetical treatment effect $\tau_0$. Likewise, let 
\begin{gather*}
Y_{i,\tau_0}^{P}(D_i,D_i^*) = (1-D_i)Y_i(0)+D_iY_i(1)+\tau_0(1-D_i)    
\end{gather*}
be the permuted outcome under the permutation procedure in \citet{henze2005checking}. Let $\mathbf{Y}_{\tau_0}^{RI}(\mathbf{D},\mathbf{D}^*)$ be a vector that stacks the outcomes permuted under randomization inference, $Y_{i,\tau_0}^{RI}(D_i,D_i^*)$, and let $\mathbf{Y}_{\tau_0}^{P}(\mathbf{D},\mathbf{D}^*)$ be  a vector that stacks the outcomes permuted under the permutation procedure, $Y_{i,\tau_0}^{P}(D_i,D_i^*)$. Let $L_{n,w}(\mathbf{Y},\mathbf{D})$ be the statistic $L_{n,w}$ calculated in the sample $(\mathbf{Y},\mathbf{D})$, $\{L_{n,w}((\mathbf{Y}^{P}_{\tau_0}(\mathbf{D},\mathbf{D}^*),\mathbf{D}^*):\mathbf{D}^*\in \mathcal{D}\}$ be the set of possible values of $L_{n,w}(\mathbf{Y}_{\tau_0}^{P}(\mathbf{D},\mathbf{D}^*),\mathbf{D}^*)$ across $\mathbf{D}^* \in \mathcal{D}$, and  $\{L_{n,w}((\mathbf{Y}^{RI}_{\tau_0}(\mathbf{D},\mathbf{D}^*),\mathbf{D}^*):\mathbf{D}^*\in \mathcal{D}\}$ be the set of possible values of $L_{n,w}(\mathbf{Y}_{\tau_0}^{RI}(\mathbf{D},\mathbf{D}^*),\mathbf{D}^*)$  across $\mathbf{D}^* \in \mathcal{D}$.
\begin{prop}
\label{prop:perm}
$\{L_{n,w}((\mathbf{Y}^{P}_{\tau_0}(\mathbf{D},\mathbf{D}^*),\mathbf{D}^*):\mathbf{D}^*\in \mathcal{D}\} = \{L_{n,w}((\mathbf{Y}^{RI}_{\tau_0}(\mathbf{D},\mathbf{D}^*),\mathbf{D}^*):\mathbf{D}^*\in \mathcal{D}\}$.
\end{prop}
\begin{proof}
Note that $L_{n,w}$ depends on differences of observed outcomes of the form $Y_i-Y_j$, where units $i$ and $j$ have the same treatment assignment. 

If $D_j^*=D_i^*$, then, under the permutation procedure,
\begin{align}
     Y_{i,\tau_0}^{P}(D_i,D_i^*)& -Y_{j,\tau_0}^{P}(D_j,D_j^*) =
    \notag
    \\
    & =(1-D_i)Y_i(0)+D_iY_i(1)+\tau_0(1-D_i)-[(1-D_j)Y_j(0)+D_jY_j(1)+\tau_0(1-D_j)],
    \notag
    \\
    & = (1-D_i)Y_i(0)+D_iY_i(1)-[(1-D_j)Y_j(0)+D_jY_j(1)]+\tau_0(D_j-D_i).
    \label{eq:lp_diff}
\end{align}
And under the randomization inference procedure,
\begin{align}
    Y_{i,\tau_0}^{RI}(D_i,D_i^*)& -Y_{j,\tau_0}^{RI}(D_j,D_j^*) =
    \notag
    \\
    & = (1-D_i)Y_i(0)+D_iY_i(1)+\tau_0(D_i^*-D_i)-[(1-D_j)Y_j(0)+D_jY_j(1)+\tau_0(D_j^*-D_j)],
    \notag
    \\
    & = (1-D_i)Y_i(0)+D_iY_i(1)-[(1-D_j)Y_j(0)+D_jY_j(1)]+\tau_0(D_j-D_i).
    \label{eq:lri_diff}
\end{align}
and, 
Then, for $D_i^*=D_j^*$, 
\[
Y_{i,\tau_0}^{RI}(D_i,D_i^*) -Y_{j,\tau_0}^{RI}(D_j,D_j^*) = Y_{i,\tau_0}^{P}(D_i,D_i^*) -Y_{j,\tau_0}^{P}(D_j,D_j^*),
\]
for $1\leq i\leq j\leq n $. Thus, $\{L_{n,w}((\mathbf{Y}^{P}_{\tau_0}(\mathbf{D},\mathbf{D}^*),\mathbf{D}^*):\mathbf{D}^*\in \mathcal{D}\} = \{L_{n,w}((\mathbf{Y}^{RI}_{\tau_0}(\mathbf{D},\mathbf{D}^*),\mathbf{D}^*):\mathbf{D}^*\in \mathcal{D}\}$.
\end{proof}

\subsection{Alternative characterization of $L_{n,\theta}$'s asymptotic distribution}
\label{sec:asympt_l_theta}

The following lemma shows the asymptotic distribution of $L_{n,\theta}$ as defined in Equation \eqref{eq:l_mine} under a moment condition. 

Let $h(y,y';\theta)=\exp\{-|y-y'|^\theta\}$ and 
\begin{gather*}
    \zeta = \sum_{d=0,1}(1-\pi_d)\left\{\EX[h(Y(d),Y'(d);\theta)h(Y(d),Y''(d);\theta)]-\EX[h(Y(d),Y'(d);\theta)]^2\right\}, 
    \\
    \mu(L_\theta) = \EX[h(Y(1),Y'(1);\theta)-h(Y(0),Y'(0);\theta)],
\end{gather*}
where $Y'(d)$ and $Y''(d)$ are independent random variables with the same distribution as $Y(d)$, $d=0,1$, and $\pi_0=1-\pi_1$.

\begin{lemma}
    \label{le:l_theta_ustat}
    Under Assumptions \ref{asm:random_assign}, \ref{asm:asm_randsample} and \ref{asm:stable_prop}, if $\EX[h(Y(d),Y'(d);\theta)^2]<\infty$, $d=0,1$, and $\zeta>0$, then
    \begin{gather*}
        \sqrt{\frac{n_0n_1}{n}}(L_{n,\theta}-\mu(L_\theta)) \rightsquigarrow \mathcal{N}(0,4\zeta).
    \end{gather*}
\end{lemma}
\begin{proof}
    Note that $h(Y(d),Y'(d);\theta)$ is a symmetric kernel of degree 2. Given Equation \eqref{eq:l_theta} and following similar steps to the derivation of Equation \eqref{eq:diff_varphi1}, one can obtain that $L_\theta = U_\theta + O(n^{-1})$ with
    \begin{gather*}
        U_\theta = \frac{1}{\binom{n_1}{2}}\sum_{j<l}^{n_1} h(Y_j(1),Y_l'(1);\theta)-\frac{1}{\binom{n_0}{2}}\sum_{j<l}^{n_0} h(Y_j(0),Y'_l(0);\theta).
    \end{gather*}
    The proof follows from Theorem 5.5.1.A in \citet{serfling2009approximation}.
\end{proof}

I now show a consistent estimator for $\zeta$ and a corresponding sample-based $t$-statistic based on $L_\theta$.
Let 
\begin{gather*}
    \widehat{\zeta}=\sum_{d=0,1}\left(1-\frac{n_d}{n}\right)\left\{R_{d}^{(3)}-\left[R_{d}^{(2)}\right]^2\right\},
    \intertext{where}
    R_{d}^{(3)}=\binom{n_d}{3}^{-1}\sum_{\substack{{j<l<k}\\D_j=D_l=D_k=d}}  h(Y_j,Y_l;\theta)h(Y_j,Y_k;\theta),
    \\
    R_{d}^{(2)}=\binom{n_d}{2}^{-1}\sum_{\substack{{j<l}\\D_j=D_l=d}} h(Y_j,Y_l;\theta)
\end{gather*}
%
The following lemma shows the consistency of $\widehat{\zeta}$ for $\zeta$.
\begin{lemma}[Consistent estimator for $\widehat{\zeta}$]
    \label{le:consistency_zeta}
     Under Assumptions \ref{asm:random_assign}, \ref{asm:asm_randsample} and \ref{asm:stable_prop}, if for $d=0,1$,
        \begin{enumerate}[(a)]
            \item \label{le:consistency_zeta_p1} $\EX[h(Y(d),Y'(d);\theta)h(Y(d),Y''(d);\theta)]<\infty$ and 
            \item \label{le:consistency_zeta_p2} $\EX[h(Y(d),Y'(d);\theta)]<\infty$, 
        \end{enumerate}
      then $\widehat{\zeta}\xrightarrow[]{P}\zeta$.
\end{lemma}
\begin{proof}
    Note that $R_{d}^{(q)}$ is a $U$-statistic with degree $q$. Under conditions \eqref{le:consistency_zeta_p1} and \eqref{le:consistency_zeta_p2}, Theorem 5.4.A in \citet{serfling2009approximation} implies that, for $d=0,1$,
    \begin{gather}
        R_{d}^{(3)} \xrightarrow[]{P} \EX[h(Y(d),Y'(d);\theta)h(Y(d),Y''(d);\theta)],
        \label{eq:limit_r3}
        \intertext{and}
        R_{d}^{(2)} \xrightarrow[]{P} \EX[h(Y(d),Y'(d);\theta)].
        \label{eq:limit_r2}
    \end{gather}
    The lemma follows from Equations \eqref{eq:limit_r3} and \eqref{eq:limit_r2}, Assumption \ref{asm:stable_prop} and the continuous mapping theorem.
\end{proof}
\begin{theorem}
    Under the conditions of Lemmas \ref{le:l_theta_ustat} and \ref{le:consistency_zeta},
    \begin{gather*}
       \frac{L_\theta-\mu(L_\theta)}{2\sqrt{\widehat{\zeta}\frac{n}{n_0n_1}}} \rightsquigarrow \mathcal{N}(0,1).
    \end{gather*}
\end{theorem}
\begin{proof}
    The theorem follows from Lemmas \ref{le:l_theta_ustat} and \ref{le:consistency_zeta} and Slutsky's theorem. 
\end{proof}

\newpage


\begin{landscape}
    
    \linespread{1}\small
    \centering
    
    \begin{ThreePartTable}
    \begin{TableNotes}[flushleft]
    \footnotesize \linespread{1}\small
    \item \footnotesize The table shows $p$-values for each of the confidence-interval Fisher randomization test (FRTCI) and the plug-in Fisher randomization test (FRTPI) in \citet{ding2016randomization}, martingale transformation test (MT Permutation) in \citet{CHUNG2021148}, and the subsampling quantile regression test (Subsampling) in \cite{chernozhukov2005subsampling}, and outcomes in \citet{augsburg2015impacts}. HH stands for household. NA stands for not applicable as the R code suggested by \citet{CHUNG2021148} did not converge. 
    \end{TableNotes}

        \begin{longtable}[t]{>{\raggedright\arraybackslash}p{4.0in}>{\raggedleft\arraybackslash}p{1.0in}>{\raggedleft\arraybackslash}p{1.0in}>{\raggedleft\arraybackslash}p{1.0in}>{\raggedleft\arraybackslash}p{1.2in}}
        \caption{\label{tab:app_ext} $p$-values for Tests for Treatment Effect Heterogeneity on Outcomes from the Microcredit Experiment in \citet{augsburg2015impacts} (Percentage Points)}\\
        \toprule
        Outcome & FRTCI & FRTPI & MT Permutation & Subsampling\\
        \midrule
        \endfirsthead
        \caption[]{$p$-values (Percentage Points) \textit{(continued)}}\\
        \toprule
        Outcome & FRTCI & FRTPI & MT Permutation & Subsampling\\
        \midrule
        \endhead
        \hline
        \insertTableNotes
        \endfoot
        \addlinespace[0.3em]
        \multicolumn{5}{l}{\textbf{Credit Outstanding at Endline}}\\
        \hspace{1em}Any loan outstanding (yes = 1) & 0.41 & 0.01 & NA & 0.0\\
        \hspace{1em}Number of loans & 10.41 & 10.01 & 10.2 & 0.0\\
        \hspace{1em}One of outstanding loans is with bank (yes = 1) & 0.01 & 0.01 & NA & 0.0\\
        \hspace{1em}At least one loan outstanding from a bank (yes = 1) & 100.01 & 0.01 & NA & 0.0\\
        \addlinespace[0.3em]
        \multicolumn{5}{l}{\textbf{Self-Employment Activities: Revenues, Assets, and Profits}}\\
        \hspace{1em}Value of all assets owned & 70.41 & 55.61 & 100.0 & 48.0\\
        \hspace{1em}Ownership of inventory (yes = 1) & 89.61 & 0.41 & NA & 0.0\\
        \hspace{1em}Average yearly revenue of main business & 77.61 & 16.41 & 0.0 & 12.8\\
        \hspace{1em}Average yearly expenses of main business & 87.21 & 32.81 & 0.0 & 31.6\\
        \hspace{1em}Average yearly profit of main business & 95.21 & 14.01 & 0.0 & 6.8\\
        \hspace{1em}Respondent owns a business (yes = 1) & 92.01 & 0.41 & 4.1 & 0.0\\
        \hspace{1em}Main business is in service (yes = 1) & 77.21 & 2.41 & NA & 0.0\\
        \hspace{1em}Main business is in agriculture/farming (yes = 1) & 73.21 & 0.41 & NA & 0.0\\
        \hspace{1em}HH has started a business (since baseline) (yes = 1) & 60.01 & 3.21 & NA & 85.2\\
        \hspace{1em}HH has closed their business (since baseline) (yes = 1) & 62.41 & 7.21 & NA & 84.0\\
        \addlinespace[0.3em]
        \multicolumn{5}{l}{\textbf{Income}}\\
        \hspace{1em}Income from self-employment received (yes = 1) & 96.81 & 2.41 & 30.6 & 0.0\\
        \hspace{1em}Amount of income from self-employment received (yearly) & 100.01 & 22.01 & 4.1 & 4.4\\
        \hspace{1em}Income from wages received (yes = 1) & 98.41 & 0.01 & 83.7 & 0.0\\
        \hspace{1em}Income from wages received (yearly) & 6.41 & 0.01 & 0.0 & 15.6\\
        \hspace{1em}Income from remittances received (yes = 1) & 74.01 & 7.61 & NA & 86.8\\
        \hspace{1em}Amount of income from remittances received (yearly) & 43.61 & 26.41 & NA & 77.6\\
        \hspace{1em}Income from government benefits received (yes = 1) & 95.21 & 5.21 & 59.2 & 0.0\\
        \hspace{1em}Amount of income from government benefits received (yearly) & 52.41 & 10.41 & NA & 72.8\\
        \addlinespace[0.3em]
        \multicolumn{5}{l}{\textbf{Consumption and Savings}}\\
        \hspace{1em}Total consumption per capita & 13.61 & 12.01 & 95.9 & 10.4\\
        \hspace{1em}Amount spent on durable consumption in last year & 98.81 & 97.61 & 0.0 & 42.0\\
        \hspace{1em}Nondurable consumption & 60.01 & 30.81 & 6.1 & 38.4\\
        \hspace{1em}Total food consumption in last week & 78.01 & 40.01 & 0.0 & 12.0\\
        \hspace{1em}Amount spent on education in last year & 28.41 & 10.41 & 65.3 & 68.4\\
        \hspace{1em}Amount spent on alcohol, cigarettes, tobacco in last week & 92.81 & 3.21 & 0.0 & 13.2\\
        \hspace{1em}Recreation & 66.01 & 30.41 & NA & 41.2\\
        \hspace{1em}Home durable good index & 0.01 & 0.01 & 4.1 & 0.0\\
        \hspace{1em}Estimated amount of savings & 82.41 & 0.81 & NA & 12.0\\
        \addlinespace[0.3em]
        \multicolumn{5}{l}{\textbf{Time Worked by Household Members}}\\
        \hspace{1em}Hours worked per HH member in a week, total & 76.81 & 74.01 & 0.0 & 11.6\\
        \hspace{1em}Hours worked per HH member in a week, business & 89.61 & 14.81 & 0.0 & 28.4\\
        \hspace{1em}Hours worked per HH member in a week, other activities & 98.41 & 8.01 & 0.0 & 0.0\\
        \hspace{1em}Hours worked per HH member in a week, total, teens & 88.01 & 2.81 & NA & 15.2\\
        \hspace{1em}Hours worked per HH member in a week, business, teens & 90.01 & 0.01 & NA & 8.8\\
        \hspace{1em}Hours worked per HH member in a week, other activities, teens & 31.61 & 26.41 & NA & 23.6\\
        \addlinespace[0.3em]
        \multicolumn{5}{l}{\textbf{Social Impacts}}\\
        \hspace{1em}Stress Score & 60.41 & 30.01 & 2.0 & 6.0\\
        \hspace{1em}Having kids in the age range 16-19 & 94.81 & 0.01 & NA & 0.0\\
        \hspace{1em}Number of kids 16-19 & 92.01 & 0.01 & NA & 0.0\\*
        \end{longtable}

    \end{ThreePartTable}

\end{landscape}

\end{document}